\documentclass[a4paper,twocolumn,amsmath,amssymb,groupedaddress,superscriptaddress,nofootinbib,accepted=2022-03-18]{quantumarticle}
\pdfoutput=1
\usepackage{mathtools}
\usepackage{braket}
\usepackage{dsfont}
\usepackage{booktabs}
\usepackage{hyperref}
\usepackage{verbatim}
\usepackage{enumitem}
\usepackage{multirow}
\usepackage{qcircuit}
\usepackage{graphicx}

\newcommand{\real}[1]{\mathfrak{Re}\{#1\}}
\newcommand{\norm}[1]{\lVert #1 \rVert}
\newcommand{\order}[1]{\mathcal{O}(#1)}
\newcommand{\R}{\mathbb{R}}
\newcommand{\id}{\mathds{1}}

\newcommand{\qwdash}[1][-1]{\ar @{.} [0,#1]}

\newcommand{\p}{\mathcal{P}}
\newcommand{\q}{\mathcal{Q}}

\newcommand{\D}{D}
\newcommand{\Dmod}{D^\ast}
\newcommand{\Dodd}{\tilde{D}}
\newcommand{\Deven}{\hat{D}}
\newcommand{\Dnoneq}{\mathring{D}} 

\newcommand{\hh}{\boldsymbol{h}}
\newcommand{\pp}{\boldsymbol{\p}}
\newcommand{\qq}{\boldsymbol{\q}}
\newcommand{\vv}{\boldsymbol{v}}
\newcommand{\xx}{\boldsymbol{x}}
\newcommand{\yy}{\boldsymbol{y}}
\newcommand{\EE}{\boldsymbol{E}}
\newcommand{\RR}{\boldsymbol{R}}
\newcommand{\HH}{\boldsymbol{H}}
\newcommand{\nnabla}{\boldsymbol{\nabla}}

\newcommand{\maxcut}{\operatorname{\textsc{MaxCut}}}

\begin{document}
\title{General parameter-shift rules for quantum gradients}
\author{David Wierichs}
\affiliation{Xanadu, Toronto, ON, M5G 2C8, Canada}
\affiliation{Institute for Theoretical Physics, University of Cologne, Germany}
\email{wierichs@thp.uni-koeln.de}
\author{Josh Izaac}
\affiliation{Xanadu, Toronto, ON, M5G 2C8, Canada}
\author{Cody Wang}
\affiliation{AWS Quantum Technologies, Seattle, Washington 98170, USA}
\author{Cedric Yen-Yu Lin}
\affiliation{AWS Quantum Technologies, Seattle, Washington 98170, USA}

\begin{abstract}
Variational quantum algorithms are ubiquitous in applications of noisy intermediate-scale
quantum computers. Due to the structure of conventional parametrized quantum gates, the
evaluated functions typically are finite Fourier series of the input parameters.
In this work, we use this fact to derive new, general parameter-shift rules for
single-parameter gates, and provide closed-form expressions to apply them.
These rules are then extended to multi-parameter quantum gates by combining them with
the stochastic parameter-shift rule.
We perform a systematic analysis of quantum resource requirements for each rule,
and show that a reduction in resources is possible for higher-order derivatives.
Using the example of the quantum approximate optimization algorithm, we show that the
generalized parameter-shift rule can reduce the number
of circuit evaluations significantly when computing derivatives with respect to
parameters that feed into many gates. Our approach additionally reproduces
reconstructions of the evaluated function up to a chosen order, leading to known
generalizations of the Rotosolve optimizer and new extensions of the quantum 
analytic descent optimization algorithm.
\end{abstract}
\maketitle

\section{Introduction}
With the advent of accessible, near-term quantum hardware, the ability to
rapidly test and prototype quantum algorithms has never been as approachable \cite{braket,arrazola2021quantum,ibmq,azure}.
However, many of the canonical quantum algorithms developed over the last
three decades remain unreachable in practice --- requiring a large number of
error corrected qubits and significant circuit depth.
As a result, a new class of quantum algorithms --- variational quantum algorithms (VQAs) \cite{benedetti2019parameterized,cerezo2021variational} --- have come
to shape the noisy intermediate-scale quantum (NISQ) era.
First rising to prominence with the introduction of the variational quantum eigensolver (VQE)
\cite{peruzzo2014variational}, they have evolved to cover topics such as optimization
\cite{qaoa},
quantum chemistry \cite{jones2019variational,qnp,grimsley2019adaptive,nakanishi2019subspace,delgado2021variational},
integer factorization \cite{anschuetz2019variational},
compilation \cite{Khatri2019quantumassisted},
quantum control \cite{parshift_0},
matrix diagonalization \cite{larose2019variational,commeau2020variational},
and variational quantum machine learning \cite{romero2017quantum,verdon2017quantum,
farhi2018classification,schuld2019quantum,parshift_1,schuld2020circuit,
grant2018hierarchical,liu2018differentiable,havlicek2019supervised,
chen2021universal,killoran2019continuous,steinbrecher2019quantum,
mari2020transfer}.

These algorithms have a common structure: a parametrized circuit is executed and a cost function is composed from expectation values measured in the resulting state.
A classical optimization routine is then used to optimize the circuit parameters by minimizing said cost function.
Initially, gradient-free optimization methods, such as Nelder-Mead and COBYLA, were common.
However, gradient-based optimization provides significant advantages, from convergence 
guarantees \cite{Sweke2020stochasticgradient} to the availability of workhorse
algorithms (e.g., stochastic gradient descent) and software tooling developed for
machine learning \cite{abadi2016tensorflow,paszke2017automatic,maclaurin2015autograd,baydin2017automatic,bergholm2018pennylane}.

The so-called parameter-shift rule \cite{parshift_0,parshift_1,parshift_2,stoch_parshift}
can be used to estimate the gradient for these optimization techniques, without additional hardware requirements and --- in contrast to na\"ive numerical methods --- without bias;
the cost function is evaluated at two shifted parameter positions, and the rescaled
difference of the results forms an unbiased estimate of the derivative.
However, this two-term parameter-shift rule is restricted to gates with two distinct
eigenvalues, potentially requiring expensive decompositions in order to compute
hardware-compatible quantum gradients \cite{crooks2019gradients}.
While various extensions to the shift rule have been discovered, they remain
restricted to gates with a particular number of distinct eigenvalues \cite{qnp,four_term_rule_kottmann}.

\begin{figure}
    \centering
    \includegraphics[width=0.475\textwidth]{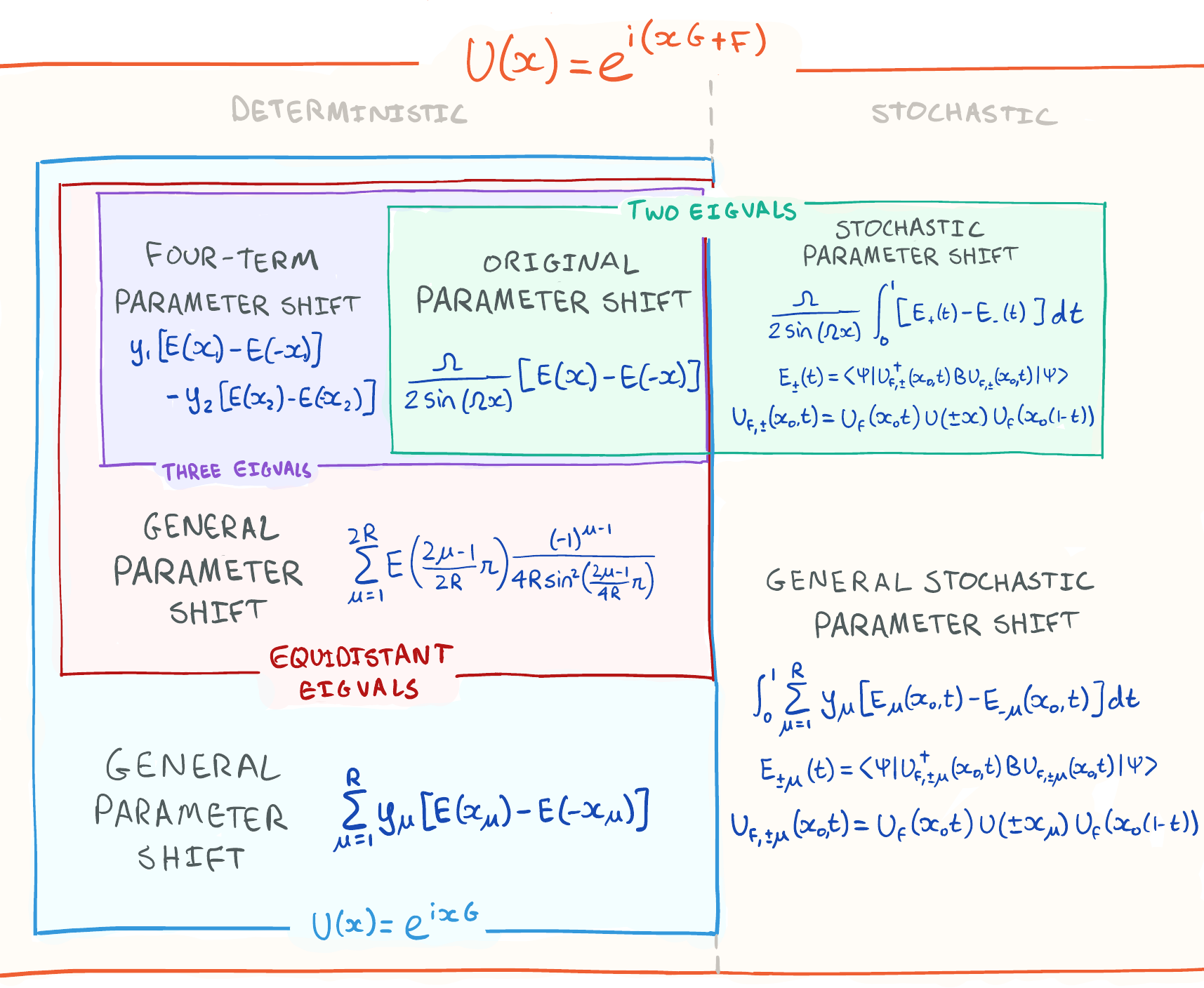}
    \caption{Overview of existing and new parameter-shift rules for first-order univariate derivatives as Venn diagram on the space of quantum gates.
    Each rule produces the analytic derivative for a set of gates, with more general rules reproducing the more specific ones.
    For gates of the form $U(x)=\exp(ixG)$ the rules are deterministic (\emph{left}) whereas more involved gates of the form $U_F=\exp(i(xG+F))$ require stochastic evaluations of shifted values (\emph{right}).
    See Sec.~\ref{sec:review_parshift} for a summary of previously known shift rules.
    The fermionic four-term shift rule in Ref.~\cite{four_term_rule_kottmann} covers the same gates as the shown four-term rule (\emph{purple}).
    }
    \label{fig:parshift_venn}
\end{figure}

In this manuscript, we use the observation that the restriction of a variational
cost function to a single parameter is a finite Fourier series
\cite{pqc_calculus,input_redundancy_PQCs,data_encoding_to_expressive_power,sequential_minimal_optimization};
as a result, the restricted cost function can be \emph{reconstructed} from
circuit evaluations at shifted positions using a discrete Fourier
transform (DFT). By analytically computing the derivatives of the Fourier
series, we extract general parameter-shift rules for arbitrary quantum gates and
provide closed-form expressions to apply them.
In the specific case of unitaries with equidistant eigenvalues, the general parameter-shift 
rule recovers known parameter-shift rules from the literature,
including the original two-term parameter-shift rule.
We then generalize our approach in two steps: first from equidistant to arbitrary 
eigenvalues of the quantum gate, and from there --- by making use of 
stochastic parameter shifts --- to more complicated unitaries like multi-parameter gates.
This enables us to cover \emph{all} practically relevant quantum gates. 
An overview of the existing parameter-shift rules and our new results is shown in Fig.~\ref{fig:parshift_venn}.

Afterwards, we perform an extensive resource analysis to compare the computational
expenses required by both the general shift rule presented here, and decomposition-based 
approaches. 
In particular, we note that evaluating the cost of gradient recipes by comparing the
number of unique executed circuits leads to fundamentally different conclusions on the
optimal differentiation technique than when comparing the total number of measurements.

Our analysis not only is fruitful for understanding the structure of
variational cost functions, but also has several practical advantages.
Firstly, second-order derivatives (such as the Hessian
\cite{Higher_order_derivatives} and the Fubini-Study metric tensor
\cite{meyer2021fisher,qng}) can be computed with fewer evaluations compared to
na\"ively iterating the two-term parameter-shift rule. 
We also show, using the example of the 
\emph{quantum approximate optimization algorithm} (QAOA),
that the generalized parameter-shift rule can reduce the number of quantum circuit 
evaluations required for ans\"atze with repeated parameters.

Finally, we generalize the \emph{quantum analytic descent} (QAD) algorithm \cite{qad}
using the reconstruction of variational cost functions discussed here.
We also reproduce the known generalizations of \emph{Rotosolve}
\cite{rotosolve,jacobi_and_anderson} from single Pauli rotations to groups of rotations
controlled by the same parameter \cite{pqc_calculus,sequential_minimal_optimization};
reconstructing functions with \emph{arbitrary} spectrum extends this algorithm even further.
Furthermore, the cost reduction for the gradient we present in the context of QAOA applies
to Rotosolve as well. 
Similarly, future improvements that reduce the cost for gradient computations might
improve the efficiency of these model-based algorithms, based on the analysis presented here.

This manuscript is structured as follows. In Sec.~\ref{sec:background}, we lay out the setting for our results by deriving the general functional form for variational cost functions, followed by a survey of existing parameter-shift rules.
In Sec.~\ref{sec:univariate_function_eq} we show how to fully reconstruct univariate variational cost functions from a finite number of evaluations assuming an equidistant frequency spectrum, and derive parameter-shift rules for arbitrary-order univariate derivatives, including a generalization of the stochastic parameter-shift rule.
In Sec.~\ref{sec:cheaper_second_order} we demonstrate how to compute second-order derivatives, in particular the Hessian and the metric tensor, more cheaply compared to existing methods.
In Sec.~\ref{sec:applications} we discuss applications, applying the new generalized
parameter-shift rules to QAOA, and using the full univariate reconstruction to extend
existing model-based optimization methods. 
We end the main text in Sec.~\ref{sec:discussion} with a discussion of our work and potential future directions.
Finally, in the appendix we summarize some technical derivations (App.~\ref{sec:derivations}), and extend the results to more general frequency spectra (App.~\ref{sec:gen}).
The general stochastic parameter-shift rule and details on quantum analytic descent can be found in Apps.~\ref{sec:derivation_stoch_parshift} and~\ref{sec:derivation_qad}. 

\emph{Related work:} In Ref.~\cite{pqc_calculus}, the functions of VQAs were considered 
as Fourier series and parameter-shift rules were derived.
Regarding the shift rules, the authors of Ref.~\cite{pqc_calculus} consider integer
eigenvalues and derive a rule with $2R+1$ evaluations for equidistant eigenvalues.
In particular, the two-term and four-term shift rules are reviewed and formulated as
special cases with \emph{fewer} evaluations than the general result presented there. 
In contrast, our work results in the exact generalization of those shift rules, 
which requires $2R$ evaluations.
Remarkably, Refs.~\cite{pqc_calculus,sequential_minimal_optimization} also propose
a generalized Rotosolve algorithm prior to its eponymous paper.

In addition, during the final stages of preparation of this work, a related work considering algebraic extensions of the parameter-shift rule appeared online \cite{algebraic_general_parshift}.
The general description of quantum expectation values in Sec.~\ref{sec:derivation_funcform} of the present work, along with its initial consequences in Sec.~\ref{sec:full_reconstruction}, are shown in Sec.~II~A of this preprint.
We present a simpler derivation and further explore the implications this description has. 
The generalization of the parameter-shift rule in Ref.~\cite{algebraic_general_parshift} is obtained by decomposing the gate generator using Cartan subalgebras, which can yield fewer shifted evaluations than decompositions of the gate itself.
In particular, decompositions into non-commuting terms, which do not lead to a gate decomposition into native quantum gates directly, can be used in this approach.

At a similar time, yet another work appeared \cite{fsim_parshift}, presenting a derivation similar to 
Sec.~\ref{sec:derivation_funcform} and parameter-shift rules for the first order derivative.
These rules are based on the ideas discussed here in Secs.~\ref{sec:full_reconstruction} and~\ref{sec:odd_reconstruction}.

\section{Background}\label{sec:background}
We start by deriving the form of a VQA cost function of a single parameter for a general single-parameter quantum gate. Then we review known parameter-shift rules and briefly discuss resource measures to compare these gradient recipes.

\subsection{Cost functions arising from quantum gates}\label{sec:derivation_funcform}
Let us first consider the expectation value for a general gate $U(x) = \exp(i x G) $, defined by a Hermitian generator $G$ and parametrized by a single parameter $x$.
Let $\ket{\psi}$ denote the quantum state that $U$ is applied to, and $B$ the measured observable\footnote{Here we consider any pure state in the Hilbert space; in the context of VQAs, $\ket{\psi}$ is the state prepared by the subcircuit prior to $U(x)$.
Similarly, $B$ includes the subcircuit following up on $U(x)$.}. The eigenvalues of $U(x)$ are given by $\left\{\exp(i\omega_j x)\right\}_{j\in [d]}$ with real-valued $\{\omega_j\}_{j\in [d]}$ where we denote $[d]\coloneqq\{1,\dots,d\}$ and have sorted the $\omega_j$ to be non-decreasing.
Thus, we have:
\begin{align}
    E(x)&\coloneqq \bra{\psi}U^\dagger(x) B U(x) \ket{\psi}\\
&=\sum_{j,k=1}^{d} \overline{\psi_j e^{i\omega_j x}} b_{jk} \psi_k e^{i\omega_k x}\\
&=\sum_{\substack{j,k=1\\j<k}}^{d} \left[\overline{\psi_j} b_{jk} \psi_k e^{i(\omega_k-\omega_j)x} \right.\\
&\hspace{1.0cm}+\left.\psi_j\overline{b_{jk} \psi_k e^{i(\omega_k-\omega_j)x}}\right]\nonumber\\
&\hspace{0.2cm}+\sum_{j=1}^{d} |\psi_j|^2 b_{jj},\nonumber
\end{align}
where we have expanded $B$ and $\ket{\psi}$ in the eigenbasis of $U$, denoted by $b_{jk}$ and $\psi_j$, respectively.

We can collect the $x$-independent part into coefficients $c_{jk}\coloneqq\overline{\psi_j}b_{jk}\psi_k$ and introduce the $R$ \emph{unique positive} differences $\{\Omega_\ell\}_{\ell\in[R]} \coloneqq \{\omega_k-\omega_j|j,k\in[d], \omega_k>\omega_j\}$.
Note that the differences are not necessarily equidistant, and that for $r=\left|\{\omega_j\}_{j\in[d]}\right|$ \emph{unique} eigenvalues of the gate generator, there are at most $R\leq \frac{r(r-1)}{2}$ unique differences. However, many quantum gates will yield $R \le r$ \emph{equidistant} differences instead; a common example for this is
\begin{align}
    G = \sum_{k=1}^\p \pm P_k
\end{align}
for commuting Pauli words $P_k$ ($P_kP_{k'} = P_{k'}P_k$), which yields the frequencies $[\p]$ and thus $R=\p$.

In the following, we implicitly assume a mapping between the two indices $j,k\in[d]$ and the frequency index $\ell\in[R]$ such that $c_\ell=c_{\ell(j,k)}$ is well-defined\footnote{That is, $\ell(j,k)=\ell(j',k')\Leftrightarrow \omega_k-\omega_j =\omega_{k'}-\omega_{j'}$.}.
We can then write the expectation value as a trigonometric polynomial (a finite-term Fourier series):
\begin{align}\label{eq:funcform}
    E(x) &=a_0+\sum_{\ell=1}^R c_\ell e^{i\Omega_\ell x}+\sum_{\ell=1}^R \overline{c_\ell} e^{-i\Omega_\ell x}\\
         &=a_0+\sum_{\ell=1}^R a_\ell \cos(\Omega_\ell x) + b_\ell\sin(\Omega_\ell x), \label{eq:E_fourier2}
\end{align}
with frequencies given by the differences $\{\Omega_\ell\}$, where we defined $c_\ell\eqqcolon \frac{1}{2}(a_\ell-ib_\ell)\;\forall \ell\in[R]$ with $a_\ell, b_\ell\in\R$, and $a_0\coloneqq \sum_j |\psi_j|^2 b_{jj}\in\R$. 

Since $E(x)$ is a finite-term Fourier series, the coefficients $\{a_\ell\}$ and $\{b_\ell\}$ can be obtained from a finite number of evaluations of $E(x)$ through a \emph{discrete Fourier transform}. This observation (and variations thereof in Sec.~\ref{sec:univariate_function_eq}) forms the core of this work: we can obtain the full functional form of $E(x)$ from a finite number of evaluations of $E(x)$, from which we can compute arbitrary order derivatives.

\subsection{Known parameter-shift rules}\label{sec:review_parshift}
\emph{Parameter-shift rules} relate derivatives of a quantum function to evaluations of the function itself at different points.
In this subsection, we survey known parameter-shift rules in the literature.

For functions of the form \eqref{eq:E_fourier2} with a single frequency $\Omega_1=\Omega$ (i.e., $G$ has two eigenvalues), the derivative can be computed via the parameter-shift rule \cite{parshift_0,parshift_1,parshift_2}
\begin{align}\label{eq:review_two_term}
    E'(0)=\frac{\Omega}{2\sin(\Omega x_1)} [E(x_1)-E(-x_1)],
\end{align}
where $x_1$ is a freely chosen shift angle from $(0, \pi)$ \footnote{The position $0$ for the derivative is chosen for convenience but the rule can be applied at any position.
To see this, note that shifting the argument of $E$ does not change its functional form.}.

This rule was generalized to gates with eigenvalues $\{-1, 0, 1\}$, which leads to $R=2$ frequencies, in Refs.~\cite{four_term_rule_kottmann,qnp} in two distinct ways.
The rule in Ref.~\cite{qnp} is an immediate generalization of the one above:
\begin{align}\label{eq:review_four_term}
    E'(0)&=y_1 [E(x_1)-E(-x_1)] \\
    &- y_2 [E(x_2)-E(-x_2)],\nonumber
\end{align}
with freely chosen shift angles $x_{1,2}$ and corresponding coefficients $y_{1,2}$, requiring four evaluations to obtain $E'(0)$.
A particularly symmetric choice of shift angles is $x_{1,2}=\pi/2\mp\pi/4$ with coefficients $y_{1,2}=\frac{\sqrt{2}\pm 1}{2\sqrt{2}}$.
In contrast, the rule in Ref.~\cite{four_term_rule_kottmann} makes use of an auxiliary gate to implement slightly altered circuits, leading to a structurally different rule:
\begin{align}\label{eq:review_four_term_Kottmann}
    E'(0)=\frac{1}{4}[E^{+}_+-E^{+}_-+E^{-}_+-E^{-}_-],
\end{align}
where $E^{\alpha}_\pm$ is the measured energy when replacing the gate $U(x)$ in question by $U(x\pm\pi/2)\exp(\mp\alpha i\frac{\pi}{4}P_0)$ and $P_0$ is the projector onto the zero-eigenspace of the generator of $U$.
Remarkably, this structure allows a reduction of the number of distinct circuit evaluations to two if the circuit and the Hamiltonian are real-valued, which is often the case for simulations of fermionic systems and forms a unique feature of this approach.
This second rule is preferable whenever this condition is fulfilled, the auxiliary gates $\exp(\pm i\frac{\pi}{4}P_0)$ are available, and simultaneously the number of distinct circuits is the relevant resource measure.

Furthermore, the two-term parameter-shift rule Eq.~\eqref{eq:review_two_term} was generalized to gates with the more complicated gate structure $U_F(x)=\exp(i(xG+F))$ via the \emph{stochastic parameter-shift rule} \cite{stoch_parshift}
\begin{align}\label{eq:review_stoch_parshift}
    E'(x_0) = \frac{\Omega}{2\sin(\Omega x_1)}\int_0^1 [E_+(t)-E_-(t)]\mathrm{d}t.
\end{align}
Here, $E_\pm(t)$ is the energy measured in the state prepared by a modified circuit that splits $U_F(x_0)$ into $U_F(tx_0)$ and $U_F((1-t)x_0)$, and interleaves these two gates with $U_{F=0}(\pm x_1)$.
See Sec.~\ref{sec:stoch_parshift} and App.~\ref{sec:derivation_stoch_parshift} for details.
The first-order parameter-shift rules summarized here and their relationship to each other is also visualized in Fig.~\ref{fig:parshift_venn}.

A parameter-shift rule for higher-order derivatives based on repeatedly applying the original rule has been proposed in Ref.~\cite{Higher_order_derivatives}.
The shift can be chosen smartly so that two function evaluations suffice to obtain the second-order derivative:
\begin{align}\label{eq:review_second_order_parshift}
    E''(0)=\frac{1}{2}[E(\pi)-E(0)],
\end{align}
which like Eq.~\eqref{eq:review_two_term} is valid for single-frequency gates.
Various expressions to compute combinations of derivatives with few evaluations were explored in Ref.~\cite{Toms_hessian}.

\subsection{Resource measures for shift rules}\label{sec:comparison_discussion}
While the original parameter-shift rule Eq.~\eqref{eq:review_two_term} provides a unique, unbiased method to estimate the derivative $E'(0)$ via evaluations of $E$ if it contains a single frequency, we will need to compare different shift rules for the general case.
To this end, we consider two resource measures.
Firstly, the number of distinct circuits that need to be evaluated to obtain all terms of a shift rule, $N_\text{eval}$.
This is a meaningful quantity on both, simulators that readily produce many measurement samples after executing each unique circuit once, as well as quantum hardware devices that are available via cloud services. In the latter case, quantum hardware devices are typically billed and queued per unique circuit, and as a result $N_\text{eval}$ often dictates both the financial and time cost.
Note that overhead due to circuit compilation and optimization scale with this quantity as well.

Secondly, we consider the overall number $N$ of measurements --- or \emph{shots} --- irrespective of the number of unique circuits they are distributed across.
To this end, we approximate the physical (one-shot) variance $\sigma^2$ of the cost function $E$ to be constant across its domain\footnote{As it is impossible in general to compute $\sigma^2$ analytically, we are forced to make this potentially very rough approximation.}. 
For an arbitrary quantity $\Delta$ computed from $\mathcal{M}$ values of $E$ via a shift rule,
\begin{align}
    \Delta=\sum_{\mu}^\mathcal{M} y_\mu E(\xx_\mu),
\end{align}
we obtain the variance for the estimate of $\Delta$ as
\begin{align}
    \varepsilon^2=\sum_{\mu}^\mathcal{M} |y_\mu|^2 \frac{\sigma^2}{N_\mu},
\end{align}
where $N_\mu$ expresses the number of shots used to measure $E(\xx_\mu)$.
For a total budget of $N$ shots, the optimal shot allocation is $N_\mu=N |y_\mu|/\norm{\yy}_1$ such that
\begin{align}
    N=\frac{\sigma^2\norm{\yy}^2_1}{\varepsilon^2}.
\end{align}
This can be understood as the number of shots needed to compute $\Delta$ to a tolerable standard deviation $\varepsilon$.

The number of shots $N$ is a meaningful quantity for simulators whose runtime scales primarily with the number of requested samples (e.g., Amazon Braket's TN1 tensor network simulator \cite{braket}),
and for actual quantum devices when artificial resource measures like pricing per unique circuit and queueing time do not play a role.

In this work we will mostly use $N_\text{eval}$ to compare the requirements of different parameter-shift rules as it is more accessible, does not rely on the assumption of constant physical variance like $N$ does, and the coefficients $\yy$ to estimate $N$ are simply not known analytically in most general cases.
For the case of equidistant frequencies and shift angles as discussed in Sec.~\ref{sec:dirichlet} we will additionally compare the number of shots $N$ in Sec.~\ref{sec:comparison_eq}.

\section{Univariate cost functions}\label{sec:univariate_function_eq}
In this section we study how a quantum cost function, which in general depends on multiple parameters, varies if only one of these parameters is changed.
The results of this section will be sufficient to evaluate the gradient as well as the diagonal of the Hessian of a quantum function.
We restrict ourselves to functions that can be written as the expectation value of an observable with respect to a state that is prepared using a unitary $U(x)=\exp(ixG)$ --- capturing the full dependence on $x$.
That is, all parameters but $x$ are fixed and the operations they control are considered as part of the prepared state and the observable.
As shown in Sec.~\ref{sec:derivation_funcform}, this yields a trigonometric polynomial, i.e.,
\begin{align}\label{eq:E_fourier}
    E(x)=a_0 + \sum_{\ell=1}^R a_{\ell}\cos(\Omega_{\ell}x)+b_{\ell}\sin(\Omega_\ell x).
\end{align}
In the following, we will assume the frequencies to be equidistant, i.e., $\Omega_\ell=\ell\Omega$, and generalize to arbitrary frequencies in App.~\ref{sec:gen}.
While it is easy to construct gate sequences that do not lead to equidistant frequencies, many conventional gates and layers of gates do yield such a regular spectrum.
The equidistant frequency case has two major advantages over the general case: we can derive closed-form parameter-shift rules (Sec.~\ref{sec:dirichlet}); and the number of circuits required for the parameter-shift rule scales much better (Sec.~\ref{sec:comparison_eq}).
 
Without loss of generality, we further restrict the frequencies to integer values, i.e., $\Omega_\ell=\ell$. For $\Omega\neq 1$, we may rescale the function argument to achieve $\Omega_\ell=\ell$ and once we reconstruct the rescaled function, the original function is available, too.

\subsection{Determining the full dependence on \texorpdfstring{$x$}{x}}\label{sec:full_reconstruction}
As we have seen, the functional form of $E(x)$ is known exactly.
We can thus determine the function by computing the $2R+1$ coefficients $\{a_\ell\}$ and $\{b_\ell\}$.
This is the well-studied problem of \emph{trigonometric interpolation} (see e.g., \cite[Chapter~X]{trig_interpolation}).

To determine $E(x)$ completely, we can simply evaluate it at $2R+1$ distinct points $x_\mu \in [-\pi, \pi)$.
We obtain a set of $2R+1$ equations
\begin{align*}
    E(x_\mu) = a_0 + \sum_{\ell=1}^R a_{\ell}\cos(\ell x_\mu)+b_{\ell}\sin(\ell x_\mu), \; \mu \in [2R]_0
\end{align*}
where we denote $[2R]_0\coloneqq\{0, 1, \dots, 2R\}$.
We can then solve these linear equations for $\{a_\ell\}$ and $\{b_\ell\}$; this process is in fact a nonuniform \emph{discrete Fourier transform (DFT)}.

A reasonable choice is $x_\mu = \frac{2 \pi \mu}{2R+1}, \mu = -R, \dots, R$, in which case the transform is the usual (uniform) DFT. For this choice, an explicit reconstruction for $E$ follows directly from \cite[Chapter~X]{trig_interpolation}; we reproduce it in App.~\ref{sec:derivation_full_reconstruction}.

\subsection{Determining the odd part of \texorpdfstring{$E(x)$}{E(x)}}\label{sec:odd_reconstruction}
It is often the case in applications that we only need to determine the odd part of $E$,
\begin{align}
    E_\text{odd}(x) &= \frac{1}{2}(E(x) - E(-x)) \label{eq:E_odd_diff} \\
    &= \sum_{\ell=1}^R b_{\ell}\sin(\ell x). \label{eq:E_odd_fourier}
\end{align}
For example, calculating odd-order derivatives of $E(x)$ at $x=0$ only requires knowledge of $E_\text{odd}(x)$, since those derivatives of the even part vanish.
Note that the reference point with respect to which $E_\text{odd}$ is odd may be chosen arbitrarily, and does not have to be $0$.

The coefficients in $E_\text{odd}$ can be determined by evaluating $E_\text{odd}$ at $R$ distinct points $x_\mu$ with $0 < x_\mu < \pi$. This gives us a system of $R$ equations
\begin{align}
    E_\text{odd}(x_\mu) = \sum_{\ell=1}^R b_{\ell}\sin(\ell x_\mu), \quad \mu\in[R]
\end{align}
which we can use to solve for the $R$ coefficients $\{b_\ell\}$.

Using Eq.~\eqref{eq:E_odd_diff} we see that each evaluation of $E_\text{odd}$ can be done with two evaluations of $E(x)$.
Thus, the odd part of $E$ can be completely determined with $2R$ evaluations of $E$, saving one evaluation compared to the general case.
Note however that the saved $E(0)$ evaluation is evaluated regardless in many applications, and may be used to recover the full reconstruction --- so, in effect, this saving does not have a significant impact\footnote{If $E(0)$ is available, we can recover the full function, allowing us to, for example, evaluate its second derivative $E''(0)$ ``for free''. However, in practice many more repetitions may be needed for reasonable accuracy. This fact was already noted in \cite{Higher_order_derivatives} for the $R=1$ case.}.

\subsection{Determining the even part of \texorpdfstring{$E(x)$}{E(x)}}\label{sec:even_reconstruction}
We might similarly want to obtain the even part of $E$,
\begin{align}
    E_\text{even}(x) &= \frac{1}{2}(E(x) + E(-x)) \label{eq:E_even_diff} \\
    &= a_0 + \sum_{\ell=1}^R a_{\ell}\cos(\ell x), \label{eq:E_even_fourier} 
\end{align}
which can be used to compute even-order derivatives of $E$.             

Determining $E_\text{even}(x)$ requires $R+1$ evaluations of $E_\text{even}$, which leads to $2R+1$ evaluations of $E$ for arbitrary frequencies. 
However, in the case where $\Omega_\ell$ are integers, $R+1$ evaluations of $E_\text{even}$ can be obtained with $2R$ evaluations of $E(x)$ by using periodicity:
\begin{align}
    E_\text{even}(0) &= E(0) \\
    E_\text{even}(x_\mu) &= \frac{1}{2}(E(x_\mu) + E(-x_\mu)), \\
    &\hspace{1cm} 0<x_\mu<\pi,\ \mu\in[R-1] \nonumber\\
    E_\text{even}(\pi) &= E(\pi).
\end{align}
Thus, in this case $2R$ evaluations of $E(x)$ suffice to determine its even part, saving one evaluation over the general case.
In contrast to the odd part, this saving genuinely reduces the required computations as $E(0)$ is also used in the cheaper computation of $\{a_\ell\}$;
therefore, if $E(0)$ is already known, we only require $2R-1$ new evaluations.

We note that even though both the odd and the even part of $E(x)$ require $2R$ evaluations, the full function can be obtained at the price of $2R+1$ evaluations.

\subsection{Explicit parameter-shift formulas}\label{sec:dirichlet}
Consider again the task of determining $E_\text{odd}$ ($E_\text{even}$) based on its value at the shifted points $\{x_\mu\}$ with $\mu\in[R]$ ($\mu\in[R]_0$).
This can be done by linearly combining elementary functions that vanish on all but one of the $\{x_\mu\}$, i.e., kernel functions, using the evaluation $E(x_\mu)$ as coefficients.
If we restrict ourselves to evenly spaced points $x_\mu = \frac{2\mu-1}{2R}\pi$ ($x_\mu = \frac{\mu}{R}\pi$), we can choose these functions to be Dirichlet kernels.
In addition to a straightforward reconstruction of the odd (even) function this delivers the \emph{general parameter-shift rules}, which we derive in App.~\ref{sec:derivation_dirichlet}:

\begin{align}\label{eq:parshift_eq}
E'(0) &= \sum_{\mu=1}^{2R} E\left(\frac{2\mu-1}{2R}\pi\right) \frac{(-1)^{\mu-1}}{4R\sin^2\left(\frac{2\mu-1}{4R}\pi\right)}, \\
E''(0) &= -E(0)\frac{2R^2+1}{6} + \sum_{\mu=1}^{2R-1} E\left(\frac{\mu\pi}{R}\right)\frac{(-1)^{\mu-1}}{2\sin^2 \left(\frac{\mu\pi}{2R}\right)} \label{eq:parshift_eq2}.
\end{align}

We remark that derivatives of higher order can be obtained in an analogous manner, and with the same function evaluations for all odd (even) orders.
Furthermore, this result reduces to the known two-term (Eq.~\eqref{eq:review_two_term}) and four-term (Eq.~\eqref{eq:review_four_term}) parameter-shift rules for $R=1$ and $R=2$, respectively, as well as the second-order derivative for $R=1$ (Eq.~\eqref{eq:review_second_order_parshift}).

We again note that the formulas above use different evaluation points for the first and second derivatives ($2R$ evaluations for each derivative). Closed-form parameter-shift rules that use $2R+1$ shared points can be obtained by differentiating the reconstruction formula Eq.~\eqref{eq:full_reconstruction}.

\subsection{Resource comparison}\label{sec:comparison_eq}
As any unitary may be compiled from (single-qubit) Pauli rotations, which satisfy the original parameter-shift rule, and CNOT gates, an alternative approach to compute $E'(0)$ is to decompose $U(x)$ into such gates and combine the derivatives based on the elementary gates.
As rotation gates about any multi-qubit Pauli word satisfy the original parameter-shift rule as well, a more coarse-grained decomposition might be possible and yield fewer evaluations for this approach.

For instance, for the $\maxcut$ QAOA ansatz\footnote{A more detailed description of the QAOA ansatz can be found in Sec.~\ref{sec:qaoa_truncation}.} on a graph $G=(\mathcal{V}, \mathcal{E})$ with vertices $\mathcal{V}$ and edges $\mathcal{E}$, one of the operations is to evolve under the problem Hamiltonian:
\begin{align}
    U_P(x) &\propto \exp\left(-i\frac{x}{2} \sum_{(a,b)\in \mathcal{E}} Z_{a} Z_{b} \right) \label{eq:qaoa_maxcut} \\
         &= \prod_{(a,b)\in \mathcal{E}} \exp \left(-i\frac{x}{2} Z_{a} Z_{b} \right). \label{eq:qaoa_maxcut_decomposed}
\end{align}
Eq.~\eqref{eq:qaoa_maxcut} treats $U_P(x)$ as a single operation with at most $M=|\mathcal{E}|$ frequencies $1, \dots, R \le M$, and we can apply the generalized parameter-shift rules of this section.
Alternatively, we could decompose $U_P(x)$ with Eq.~\eqref{eq:qaoa_maxcut_decomposed}, apply the two-term parameter-shift rule to each $R_{ZZ}$ rotation, and sum up the contributions using the chain rule.

\subsubsection{Number of unique circuits}
If there are $\p$ gates that depend on $x$ in the decomposition, this approach requires $2\p$ unique circuit evaluations; as a result, the general parameter-shift rule is cheaper if $R<\p$.
The evaluations used in the decomposition-based approach cannot be expressed by $E$ directly because the parameter is shifted only in one of the $\p$ gates per evaluation, which makes the general parameter-shift rule more convenient and may reduce compilation overhead for quantum hardware, and the number of operations on simulators.

In order to compute $E''(0)$ via the decomposition, we need to obtain and sum the full Hessian of all elementary gates that depend on $x$ (see App.~\ref{sec:derivation_decomp_coefficient_norm}), which requires $2\p^2-\p+1$ evaluations, including $E(0)$, and thus is significantly more expensive than the $2R$ evaluations for the general parameter-shift rule.

While the derivatives can be calculated from the functional form of $E_\text{odd}$ or $E_\text{even}$, the converse is not true for $R>1$, i.e., the full functional dependence on $x$ cannot be extracted from the first and second derivative alone.
Therefore, the decomposition-based approach would demand a full multivariate reconstruction for all $\p$ parametrized elementary gates to obtain this dependence, requiring $\order{2^{\p}}$ evaluations.
The approach shown here allows us to compute the dependence in $2R+1$ evaluations and thus is the only method for which the univariate reconstruction is viable.

\subsubsection{Number of shots}
For equidistant evaluation points, we explicitly know the coefficients of the first and second-order shift rule given in Eqs.~(\ref{eq:parshift_eq}, \ref{eq:parshift_eq2}), and thus can compare the variance of the derivatives in the context and under the assumptions of Sec.~\ref{sec:comparison_discussion}.

The coefficients satisfy (see App.~\ref{sec:derivation_parshift_coefficient_norm})
\begin{align*}
    \sum_{\mu=1}^{2R}\left(4R\sin^2\left(\frac{2\mu-1}{4R}\pi\right)\right)^{-1}&=R\\
    \frac{2R^2+1}{6}+\sum_{\mu=1}^{2R-1}\left(2\sin^2\left(\frac{\mu\pi}{2R}\right)\right)^{-1}&=R^2.
\end{align*}
This means that the variance-minimizing shot allocation requires a shot budget of
\begin{align}
    N_\text{genPS, 1}&=\frac{\sigma^2R^2}{\varepsilon^2}\label{eq:shots_genpar1}\\
    N_\text{genPS, 2}&=\frac{\sigma^2R^4}{\varepsilon^2}\label{eq:shots_genpar2}
\end{align}
using the generalized parameter-shift rule for the first and second derivative, respectively.

Assuming integer-valued frequencies in the cost function typically means, in the decomposition-based approach, that $x$ enters the elementary gates without any additional prefactors\footnote{Of course, one can construct less efficient decompositions that do not satisfy this rule of thumb.}.
Thus, optimally all evaluations for the first-order derivative rule are performed with the same portion of shots; whereas the second-order derivative requires an adapted shot allocation which, in particular, measures $E(0)$ with high precision as it enters $E''(0)$ with the prefactor $\p/2$.
This yields (see App.~\ref{sec:derivation_decomp_coefficient_norm})
\begin{align}
    N_\text{decomp, 1}&=\frac{\sigma^2 \p^2}{\varepsilon^2}\\
    N_\text{decomp, 2}&=\frac{\sigma^2 \p^4}{\varepsilon^2}.
\end{align}
Comparing with $N_\text{genPS, 1}$ and $N_\text{genPS, 2}$ above, we see that the shot budgets are equal at $\p=R$.
That is, for both the first and second derivative, the general parameter-shift rule does not show lower shot requirements in general, in contrast to the previous analysis that showed a significantly smaller number of unique circuits for the second derivative.
This shows that the comparison of recipes for gradients and higher-order derivatives crucially depends on the chosen resource measure. 
In specific cases we may be able to give tighter upper bounds on $R$ so that $R < \p$ (see Sec.~\ref{sec:qaoa_truncation}) and the general shift rule becomes favourable regarding the shot count as well.

\subsection{General stochastic parameter-shift rule}\label{sec:stoch_parshift}
Next, we will apply the \emph{stochastic parameter-shift rule} to our general shift rule.
For this section we do \emph{not} assume the frequencies to be equidistant but address  arbitrary spectra directly. Additionally we make the reference point $x_0$ at which the derivative is computed explicit.

In Ref.~\cite{stoch_parshift}, the authors derive the stochastic parameter-shift rule for gates of the form
\begin{align}
    U_F(x) = \exp(i(xG+F))
\end{align}
where $G$ is a Hermitian operator with eigenvalues $\pm 1$ (so that $G^2=\id$), e.g., a Pauli word.
$F$ is any other Hermitian operator, which may not necessarily commute with $G$\footnote{If $GF=FG$, the exponential may be split into $\exp(ixG)$ and $\exp(iF)$ and we are back at the situation $\exp(ixG)$.}.
Key to the derivation of the stochastic rule is an identity relating the derivative of the quantum channel $\mathcal{U}_F(x)[\rho]=U_F^\dagger(x) \rho U_F(x)$ to the derivative of the generator channel $\mathcal{G}(x)[\rho]=i[(xG+F), \rho]$.
We may extend this directly to the general parameter-shift rule for the case when $G^2 = \id$ is no longer satisfied (see App.~\ref{sec:derivation_stoch_parshift} for the derivation):

\begin{align}\label{eq:stoch_parshift}
    E'(x_0)&=\int_0^1 \sum_{\mu=1}^R y_\mu [E_\mu(x_0,t)-E_{-\mu}(x_0, t)]\mathrm{d}t\\
    E_{\pm \mu}(x_0, t)&\coloneqq \langle B\rangle_{U_F(tx_0)U(\pm x_\mu)U_F((1-t)x_0)\ket{\psi}}.\nonumber
\end{align}
The integration is implemented in practice by sampling values for $t$ for each measurement of $E_{\mu}(x_0, t)$ and $E_{-\mu}(x_0, t)$.

The stochastic parameter-shift rule in combination with the generalized shift rule in Eq.~\eqref{eq:parshift_eq} allows for the differentiation of any unitary with equidistant frequencies.
As $F$ in $U_F(x)$ above is allowed to contain terms that depend on other variational parameters, this includes multi-parameter gates in particular.
Furthermore, combining Eq.~\eqref{eq:stoch_parshift} with the generalized shift rule for arbitrary frequencies in Eq.~\eqref{eq:parshift_gen} allows us to compute the derivative of \emph{any} quantum gate as long as the frequencies of $U_{F=0}(x)$ are known.
We thus obtain an improved rule for $U_{F\neq 0}(x)$ over the original stochastic shift rule whenever the generalized shift rule is beneficial for $U(x)=U_{F=0}(x)$, compared to the decomposition-based approach.

\section{Second-order derivatives}\label{sec:cheaper_second_order}
As noted in Sec.~\ref{sec:even_reconstruction}, higher-order derivatives of univariate functions are easily computed using the even or odd part of the function. In the following sections, we will extend our discussion to multivariate functions $E(\xx)$, where derivatives may be taken with respect to different variables.
Each single parameter dependence is assumed to be of the form Eq.~\eqref{eq:funcform}, with equidistant (and by rescaling integer-valued) frequencies $\{\Omega_\ell^{(k)}\}_{\ell\in[R_k]}=[R_k]$ for the $k$th parameter. We may collect the numbers of frequencies in a vector $(\RR)_k=R_k$.
It will again be useful in the following to make the reference point $\xx_0$, at which these derivatives are computed, explicit.

\subsection{Diagonal shift rule for the Hessian}\label{sec:cheaper_hessian}
Here we show how to compute the Hessian $H$ of a multivariate function $E(\xx)$ at some reference point $\xx_0$ using the Fourier series representation of $E$.
We allow for single-parameter gates $U(x)=\exp(ixG)$ with equidistant frequencies and will use fewer evaluations of $E$ than known schemes.
An indication that this may be possible for gates with two eigenvalues was made in \cite[Eq.~(37)]{Toms_hessian}.

First, for the $k$th diagonal entry $H_{kk}=\partial^2_k E(\xx_0)$ of the Hessian, we previously noted in Sec.~\ref{sec:even_reconstruction} that $2R_k$ evaluations are sufficient as it is the second derivative of a univariate restriction of $E$.
Recall that one of the $2R_k$ evaluations is $E(\xx_0)$; we can reuse this evaluation for all diagonal entries of $H$, and thus require $1+\sum_{k=1}^n (2R_k-1)=2\norm{\RR}_1-n+1$ evaluations for the full diagonal.
Further, if we compute the Hessian diagonal $(\nnabla^{\odot 2}E)_k\coloneqq \partial_k^2 E$ in addition to the gradient, we may reuse the $2\norm{\RR}_1$ evaluations computed for the gradient, only requiring a single additional function value, namely $E(\xx_0)$.
In this case, we do not make use of the periodicity $E(\xx_0+\pi\vv_k)=E(\xx_0-\pi\vv_k)$, where $\vv_k$ is the $k$th canonical basis vector, because this shift is not used in the gradient evaluation (see Sec.~\ref{sec:odd_reconstruction}).

Next, for an off-diagonal entry $H_{km}=\partial_k\partial_m E(\xx_0)$, consider the \emph{univariate} trigonometric function that shifts the two parameters $x_k$ and $x_m$ \emph{simultaneously}:
\begin{align}\label{eq:simul_shift}
    E^{(km)}(x) \coloneqq E(\xx_0+x\vv_{k,m}),
\end{align}
where we abbreviated $\vv_{k,m} \coloneqq \vv_k+\vv_m$.
We show in App.~\ref{sec:derivation_hessian_rule} that $E^{(km)}$ again is a Fourier series of $x$ with $R_{km}=R_k+R_m$ equidistant frequencies. 
This means that we can compute ${E^{(km)}}''(0)$ via Eq.~\eqref{eq:parshift_eq2} with $R=R_{km}$, using $2R_{km}-1$ evaluations of $E$ (as we may reuse $E(\xx_0)$ from the diagonal computation).
Note that
\begin{align}
    \left. \frac{\mathrm{d}^2}{\mathrm{d}x^2} E^{(km)}(x)\right|_{x=0} = H_{kk}+H_{mm}+2H_{km},
\end{align}
and that we have already computed the diagonal entries. We thus may obtain $H_{km}$ via the \emph{diagonal parameter-shift rule}
\begin{align}\label{eq:hessian_parshift}
    H_{km} = \frac{1}{2}\left({E^{(km)}}''(0)-H_{kk}-H_{mm}\right).
\end{align}

\begin{figure}
    \centering
    \includegraphics[width=0.45\textwidth]{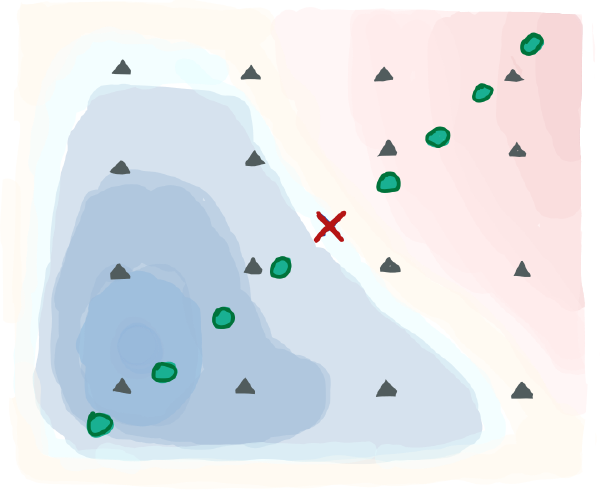}
    \caption{Visual representation of two approaches to compute a Hessian entry $H_{km}$ at the position $\xx_0$ (\emph{red cross}).
    The parameters $x_k$ and $x_m$ lie on the coordinate axes and the heatmap displays the cost function $E(\xx)$.
    We may either combine the general shift rule for $x_k$ and $x_m$ (\emph{grey triangles}) or compute the univariate derivative ${E^{(km)}}''(0)$ and extract $H_{km}$ via Eq.~\eqref{eq:hessian_parshift} (\emph{green circles}).
    }
    \label{fig:hessian_parshift}
\end{figure}

In Fig.~\ref{fig:hessian_parshift}, we visually compare the computation of $H_{km}$ via the diagonal shift rule to the chained application of univariate parameter-shift rules for $x_k$ and $x_m$.

As an example, consider the case when $R_k=R_m=1$ (e.g., where all parametrized gates are of the form $\exp(ix_k G_k/2)$ with $G_k^2=\id$). By setting $R=2$ in Eq.~\eqref{eq:parshift_eq2}, we obtain the explicit formula for ${E^{(km)}}''(0)$,
\begin{align}
    {E^{(km)}}''(0) &= -\frac{3}{2}E(\xx_0) -\frac{1}{2}E(\xx_0+\pi \vv_{k,m})\\
    &+ E\left(\xx_0+\frac{\pi}{2}\vv_{k,m}\right) + E\left(\xx_0-\frac{\pi}{2}\vv_{k,m}\right)\nonumber 
\end{align}
which can be combined with Eq.~\eqref{eq:hessian_parshift} to give an explicit formula for the Hessian. This formula (for $R_k=R_m=1$) was already discovered in \cite[Eq.~(37)]{Toms_hessian}.

The computation of $H_{km}$ along the main diagonal in the $x_k$-$x_m$-plane can be modified by making use of the second diagonal as well:
define $\overline{\vv}_{k,m}\coloneqq \vv_k-\vv_m$ and $\overline{E}^{(km)}(x)\coloneqq E(\xx_0+x\overline{\vv}_{k,m})$, and compute
\begin{align}
    \left.\frac{\text{d}^2}{\text{d}x^2}\overline{E}^{(km)}(x)\right|_{x=0}&=H_{kk}+H_{mm}-2H_{km},\\
H_{km} &= \frac{1}{4}\left({E^{(km)}}''(0)-{\overline{E}^{(km)}}''(0)\right).\nonumber
\end{align}
This means we can replace the dependence on the diagonal elements $H_{kk}$ and $H_{mm}$ by another univariate second-order derivative on the second diagonal.
We will not analyze the resources required by this method in detail but note that for many applications it forms a compromise between the two approaches shown in Fig.~\ref{fig:hessian_parshift}.

We note that an idea similar to the ones presented here can be used for higher-order derivatives, but possibly requires more than one additional univariate reconstruction per derivative.

\subsection{Resource comparison}\label{sec:comparison_hess}
For the Hessian computation, we will again look at the number of unique circuit evaluations $N_\text{eval}$ and the number of shots $N$, as introduced in Sec.~\ref{sec:comparison_discussion}.

\subsubsection{Number of unique circuits}

\begin{table*}
\centering
\begin{tabular}{llll}
    Quantity & Decomposition & Gen. shift rule, equidistant & Gen. shift rule \\
   \hline
    $E(\xx_0)$ & $1$ & $1$ & $1$\\
    $\partial_k E(\xx_0)$ & $2\p_k$ & $2R_k$ & $2R_k$\\
    $\nnabla E(\xx_0)$ & $2\norm{\pp}_1$ & $2\norm{\RR}_1$ & $2\norm{\RR}_1$\\
    $\partial_k^2 E(\xx_0)$ & $2\p_k^2-\p_k+1$ & $2R_k$ & $2R_k+1$\\
    $\nnabla^{\odot 2} E(\xx_0)$ & $2\norm{\pp}_2^2-\norm{\pp}_1+1$ & $2\norm{\RR}_1-n+1$ & $2\norm{\RR}_1+1$\\
    $\partial_k\partial_m E(\xx_0)$ & $4\p_k\p_m$ & $2(R_k+R_m)-1 ^{(\ast)}$ & $4R_kR_m+2R_k+2R_m-4 ^{(\ast)}$\\
    \multirow{2}{*}{$\nnabla^{\otimes 2} E(\xx_0)$} &
    \multirow{2}{*}{$2\norm{\pp}_1^2-\norm{\pp}_1+1$} & 
    \multirow{2}{*}{$2n\norm{\RR}_1 -\frac{1}{2}(n^2+n-2)$} & $2\left(\norm{\RR}_1^2-\norm{\RR}_2^2+n\norm{\RR}_1\right)$ \\
    &  &  & $\hspace{2.5cm} -2n(n-1)+1$ \\
    \hline
    $\partial_k E(\xx_0)$ \& $\partial_k^2 E(\xx_0)$ & $2\p_k^2+1$ & $2R_k+1$ & $2R_k+1$ \\
    $\nnabla E(\xx_0)$ \& $\nnabla^{\odot2} E(\xx_0)$ & $2\norm{\pp}_2^2+1$ & $2\norm{\RR}_1+1$ & $2\norm{\RR}_1+1$ \\
    \multirow{2}{*}{$\nnabla E(\xx_0)$ \& $\nnabla^{\otimes 2} E(\xx_0)$} &
    \multirow{2}{*}{$2\norm{\pp}_1^2+1$} & 
    \multirow{2}{*}{$2n\norm{\RR}_1 -\frac{1}{2}(n^2-n-2)$} &
    $2\left(\norm{\RR}_1^2-\norm{\RR}_2^2+n\norm{\RR}_1\right)$\\
    &  &  & $\hspace{2.5cm} -2n(n-1)+1$ \\
\end{tabular}
\caption{Number of distinct circuit evaluations $N_\text{eval}$ for measuring combinations of derivatives of a parametrized expectation value function $E$ at parameter position $\xx_0$.
The compared approaches include decomposition of the unitaries together with the original parameter-shift rule (\emph{left}), and the generalized parameter-shift rule Eq.~\eqref{eq:parshift_eq} together with the diagonal shift rule for the Hessian in Eq.~\eqref{eq:hessian_parshift}.
The requirements for the latter differ significantly for equidistant (\emph{center}) and arbitrary frequencies (\emph{right}, see App.~\ref{sec:multivariate_gen}).
A third approach is to repeat the general parameter-shift rule, the cost of which can be read off by replacing $\pp$ by $\RR$ in the left column.
Here, $n$ is the number of parameters in the circuit, $\p_k$ is the number of elementary gates with two eigenvalues in the decomposition of the $k$th parametrized unitary, and $R_k$ denotes the number of frequencies for the $k$th parameter.
The asterisk ${}^{(\ast)}$ indicates that the derivatives $\partial^2_k E$ and $\partial^2_m E$ need to be known in order to obtain the mixed derivative at the shown price (see main text).
The evaluation numbers take savings into account that are based on using evaluated energies for multiple derivative quantities; hence, they are not additive in general.
\label{tab:quantity_cost}}
\end{table*}

In Tab.~\ref{tab:quantity_cost}, we summarize the number of distinct circuit evaluations required to compute several combinations of derivatives of $E(\xx)$, either by decomposing the gate or by using the general parameter-shift rule together with the diagonal shift rule for the Hessian.
We also include the generalized case of non-equidistant frequencies covered in App.~\ref{sec:multivariate_gen} for completeness.
To obtain the cost for the repeated general shift rule, i.e., without the diagonal shift rule for the Hessian or decomposition, simply replace $\pp$ by $\RR$ in the left column.

For equidistant frequencies, the diagonal shift rule for $H_{km}$ requires $2(R_k+R_m)-1$ evaluations, assuming the diagonal and thus $E(\xx_0)$ to be known already.
Like the gradient, $H_{km}$ may instead be computed by decomposing $U_k(x_k)$ and $U_m(x_m)$ into $\p_k$ and $\p_m$ elementary gates, respectively, and repeating the parameter-shift rule twice \cite{Higher_order_derivatives, metric_tensor_proj_meas}.
All combinations of parameter shifts are required, leading to $4\p_k\p_m$ evaluations.
Finally, as a third option, one may repeat the general parameter-shift rule in Eq.~\eqref{eq:parshift_eq} twice, leading to $4R_kR_m$ evaluations\footnote{These $4R_kR_m$ shifted evaluations are \emph{not} simultaneous shifts in both directions of the form Eq.~\eqref{eq:simul_shift}.}.

The repeated general shift rule requires strictly more circuit evaluations than the diagonal shift rule, since
\begin{align}
    2\norm{\RR}_1^2-\norm{\RR}_1+1>2n\norm{\RR}_1-\frac{1}{2}(n^2+n-2).
\end{align}
Similar to the discussion for the scaling of gradient computations, the optimal approach depends on $R_{k,m}$ and $\p_{k,m}$, but $\p$ and $R$ often have a linear relation so that the diagonal shift rule will be
significantly cheaper for many cost functions than decomposing the unitaries.

\subsubsection{Number of shots}
Next we compare the numbers of measurements required to reach a precision $\varepsilon$. While the approach via repeated shift rules uses distinct circuit evaluations for each Hessian entry, the diagonal shift rule in Eq.~\eqref{eq:hessian_parshift} reuses entries of the Hessian and thus correlates the optimal shot allocations and the statistical errors of the Hessian entries. We therefore consider an error measure on the full Hessian matrix instead of a single entry, namely the root mean square of the Frobenius norm of the difference between the true and the estimated Hessian.
This norm is computed in App.~\ref{sec:derivation_coefficient_norms_hessian} for the three presented approaches, and we conclude the number of shots required to achieve a norm of $\varepsilon$ to be
\begin{align}
    N_\text{diag}&=\frac{\sigma^2}{2\varepsilon^2}\Big[\bigl(\sqrt{n+1}+n-2\bigr)\norm{\RR}_2^2+\norm{\RR}_1^2\Big]^2\\
    N_\text{genPS} &= \frac{\sigma^2}{2\varepsilon^2}\Big[\bigl(\sqrt{2}-1\bigr)\norm{\RR}_2^2+\norm{\RR}_1^2\Big]^2\\
    N_\text{decomp} &= \frac{\sigma^2}{2\varepsilon^2}\Big[\bigl(\sqrt{2}-1\bigr)\norm{\pp}_2^2+\norm{\pp}_1^2\Big]^2
\end{align}
In general, the diagonal shift rule for the Hessian is significantly less efficient than the repeated execution of the general parameter-shift rule if the shot count is the relevant resource measure.
This is in sharp contrast to the number of unique circuits, which is strictly smaller for the diagonal shift rule.
We note that the two resource measures yield \emph{incompatible} recommendations for the computation of the Hessian.
The overhead of the diagonal shift rule reduces to a (to leading order in $n$) constant prefactor if $R_k=R$ for all $k\in[n]$:
in this case, we know $\norm{\RR}_1=n=\norm{\RR}_2^2$ and therefore
\begin{align}
    \frac{N_\text{diag}}{N_\text{genPS}}=\frac{2n+\sqrt{n+1}-2}{n+\sqrt{2}-1}\underset{n\to\infty}{\longrightarrow} 2.
\end{align}

\subsection{Metric tensor}\label{sec:metric_tensor}
The Fubini-Study metric tensor $\mathcal{F}$ is the natural metric on the manifold of (parametrized) quantum states, and the key ingredient in quantum natural gradient descent \cite{qng}.
The component of the metric belonging to the parameters $x_k$ and $x_m$ can be written as
\begin{align}
    \mathcal{F}_{km}(\xx_0) =& \real{\braket{\partial_k\psi(\xx)|\partial_m\psi(\xx)}}\,\Big|_{\xx = \xx_0}\\
    &-\braket{\partial_k\psi(\xx)|\psi(\xx)}\braket{\psi(\xx)|\partial_m\psi(\xx)}\,\Big|_{\xx = \xx_0},\nonumber
\end{align}
or, alternatively, as a Hessian \cite{Higher_order_derivatives}:
\begin{align}\label{eq:metric_as_hessian}
\mathcal{F}_{km}(\xx_0) &= -\frac{1}{2} \partial_k \partial_m |\!\braket{\psi(\xx) | \psi(\xx_0)}\!|^2 \,\Big|_{\xx = \xx_0}\nonumber\\
&\eqqcolon \partial_k\partial_m f(\xx_0).
\end{align}

It follows that we can compute the metric using the same method as for the Hessian, with $f(\xx)$ as the cost function.
We know the value of $f$ without shift as
\begin{align}
    f(\xx_0)=-\frac{1}{2}|\!\braket{\psi(\xx_0)|\psi(\xx_0)}\!|^2=-\frac{1}{2}.
\end{align}
The values with shifted argument can be calculated as the probability of the zero bitstring $\mathbf{0}$ when measuring the state $V^\dagger(\xx) V(\xx_0) \ket{\mathbf{0}}$ in the computational basis, which requires circuits with up to doubled depth compared to the original circuit $V(\xx)$.
Alternatively, we may use a Hadamard test to implement $f$, requiring an auxiliary qubit, two operations controlled by that qubit as well as a measurement on it, but only halved depth on average (see App.~\ref{sec:derivation_Hadamard_test}).
With either of these methods, the terms for the shift rule in Eq.~\eqref{eq:hessian_parshift} and thus the metric tensor can be computed via the parameter-shift rule.

The metric can also be computed analytically without parameter shifts via a \emph{linear combination of unitaries (LCU)} \cite{variational_ITE,variational_sim_active_error_minimization}, which also employs Hadamard tests.
As it uses the generator as an operation in the circuit, any non-unitary generator needs to be decomposed into Pauli words for this method to be available on quantum hardware, similar to a gate decomposition.
Afterwards, this method uses one circuit evaluation per pair of Pauli words from the $k$th and $m$th generator to compute the entry $\mathcal{F}_{km}$.
A modification of all approaches that use a Hadamard test is possible by replacing it with projective measurements \cite{metric_tensor_proj_meas}.

Metric entries that belong to operations that commute \emph{within the circuit}\footnote{For example, operations on distinct wires commute in general but not necessarily within the circuit if entangling operations are carried out between them.} can be computed block-wise without any auxiliary qubits, additional operations or deeper circuits \cite{qng}.
For a given block, we execute the subcircuit $V_1$ prior to the group of mutually commuting gates and measure the covariance matrix of the generators $\{G_k\}$ of these gates:
\begin{align}
    \mathcal{F}_{km}&=\bra{\mathbf{0}}V_1^\dagger G_kG_m V_1\ket{\mathbf{0}}\\
    &-\bra{\mathbf{0}}V_1^\dagger G_k V_1\ket{\mathbf{0}} \bra{\mathbf{0}}V_1^\dagger G_m V_1\ket{\mathbf{0}}.\nonumber
\end{align}
By grouping the measurement bases of all $\{G_kG_m\}$ and $\{G_k\}$ of the block, the covariance matrix can typically be measured with only a few unique circuit evaluations\footnote{For a layer of simultaneous single-qubit rotations on all $N$ qubits, even a single measurement basis is sufficient for the corresponding $N\times N$ block.}, making this method the best choice for the block-diagonal.
One may then either use the result as an approximation to the full metric tensor, or use one of the other methods to compute the off-block-diagonal entries;
the approximation has been shown to work well for some circuit structures \cite{qng}, but not for others \cite{avoiding}.
The methods to obtain the metric tensor and their resource requirements are shown in Tab.~\ref{tab:metric_tensor_methods}.

\begin{table*}
\centering
\begin{tabular}{lccccc}
    & \multicolumn{2}{c}{Parameter shift rule} & LCU & Covariance\\
    & \hspace{1.4cm}Overlap & Hadamard & \\
    \hline
    Aux. qubits & \hspace{1.8cm}$0$ & $1$ & $1$ & $0$ \\
    off-block-diag. & \hspace{1.8cm}$\checkmark$ & $\checkmark$ & $\checkmark$ &  \\
    Depth (avg) & \hspace{1.8cm}$\sim \frac{4}{3}D_V$ & $\sim \frac{2}{3}D_V$ & $\sim \frac{2}{3}D_V$ & $\frac{2}{3}D_V$\\
    Depth (max) & \hspace{1.8cm}$2D_V$ & $\sim D_V$ & $\sim D_V$ & $D_V$ \\
    \hline
    $N_\text{eval} (\mathcal{F}_{kk})$ & 
    \multicolumn{2}{l}{$
    \begin{cases}
        2R_k-1 \\
        2R_k \\
    \end{cases}
    $} &
    $\q_k\leq \frac{1}{2}(\p_k^2-\p_k)$ & 
    $\overline{\p}_k\leq \p_k$ \\
    $N_\text{eval} (\mathcal{F}_{km})$ & 
    \multicolumn{2}{l}{$
    \begin{cases}
        2(R_k+R_m)-1 \\
        2(2R_kR_m+R_k+R_m-2)\\
    \end{cases}
    $} &
    $\p_k\p_m$ &
    $\overline{\p}_{km}\leq \p_k\p_m$ \\
    $N_\text{eval} (\mathcal{F})$ & 
    \multicolumn{2}{l}{$
    \begin{cases}
        2n\norm{\RR}_1-\frac{1}{2}(n^2+n)\\
        2\left(\norm{\RR}_1^2-\norm{\RR}_2^2+n(\norm{\RR}_1-n+1)\right)\\
    \end{cases}
    $} &
    $\frac{1}{2}\left(\norm{\pp}_1^2-\norm{\pp}_2^2\right)+\norm{\qq}_1$ & ---\\
\end{tabular}
\caption{Quantum hardware-ready methods to compute the Fubini-Study metric tensor and their resource requirements. The cost function $f(\xx)$ (see Eq.~\eqref{eq:metric_as_hessian}) for the parameter-shift rule can be implemented with increased depth by applying the adjoint of the original circuit to directly realize the overlap (\emph{left}) or with an auxiliary qubit and Hadamard tests (\emph{center left}, App.~\ref{sec:derivation_Hadamard_test}).
The LCU method (\emph{center right}) is based on Hadamard tests as well and both these methods can spare the auxiliary qubit and instead employ projective measurements \cite{metric_tensor_proj_meas}.
The cheapest method is via measurements of the covariance of generators (\emph{right}) but it can only be used for the block-diagonal of the tensor, i.e., not for all $\mathcal{F}_{km}$.
We denote the depth of the original circuit $V$ by $D_V$ and the number of Pauli words in the decomposition of $G_k$ and its square with $\p_k$ and $\q_k$, respectively.
The $\p_k$ Pauli words of $G_k$ can be grouped into $\overline{\p}_k$ groups of pairwise commuting words; the number of groups of pairwise commuting Pauli words in the product $G_kG_m$ similarly is $\overline{\p}_{km}$.
For the covariance-based approach, we overestimate the number of required circuits, as typically many of the measurement bases of the entries in the same block will be compatible.
The number of unique circuits to be evaluated for a diagonal element $\mathcal{F}_{kk}$, an off-diagonal element $\mathcal{F}_{km}$, and the full tensor $\mathcal{F}$ is given in terms of the number of frequencies $R_k$ and of $\q_k$, $\p_k$ $\overline{\p}_k$ and $\overline{\p}_{km}$.
The entries for $N_\text{eval}$ in the first and second row of the braces refer to equidistant (main text) and arbitrary frequencies (see App.~\ref{sec:multivariate_gen}), respectively.}
\label{tab:metric_tensor_methods}
\end{table*}

Since we run a different circuit for the metric tensor than for the cost function itself, the $2 R_k-1$ evaluations at shifted positions needed for the $k$th diagonal entry cannot reuse any prior circuit evaluations, as is the case for the cost function Hessian.
Consequentially, the natural gradient of a (single term) expectation value function $E$,

\begin{align}
\nnabla\!_\text{n} \ E(\xx) := \mathcal{F}^{-1}(\xx) \nnabla E(\xx),
\end{align}
with $\nnabla E$ referring to the Euclidean gradient, requires more circuit evaluations than its Hessian and gradient together.

However, the utility of the metric tensor becomes apparent upon observing that it depends solely on the \emph{ansatz}, and not the observable being measured.
This means that if a cost function has multiple terms, like in VQEs, the metric only needs to be computed once per epoch, rather than once per term, as is the case of the cost function Hessian.
Therefore, an epoch of quantum natural gradient descent can be cheaper for such cost functions than an epoch of optimizers using the Hessian of the cost function.
In addition, the block-diagonal of the metric tensor can be obtained with few circuit evaluations per block for conventional gates without any further requirements and with reduced average circuit depth.

\section{Applications}\label{sec:applications}
In this section, we will present QAOA as concrete application for our general parameter-shift rule, which reduces the required resources significantly when computing derivatives.
Afterwards, we use the approach of trigonometric interpolation to generalize the Rotosolve
algorithm. This makes it applicable to arbitrary quantum gates with equidistant
frequencies, which reproduces the results in
Refs.~\cite{pqc_calculus,sequential_minimal_optimization}, and extends them further to more
general frequency spectra.
In addition, we make quantum analytic descent (QAD) available for arbitrary quantum gates with equidistant frequencies, which previously required a higher-dimensional
Fourier reconstruction and thus was infeasible.

\subsection{QAOA and Hamiltonian time evolution}\label{sec:qaoa_truncation}
In Eq.~\eqref{eq:parshift_eq} we presented a generalized parameter-shift rule that makes use of $2R$ function evaluations for $R$ frequencies in $E$.
A particular example for single-parameter unitaries with many frequencies are layers of single- or two-qubit rotation gates, as can be found e.g., in QAOA circuits or digitized Hamiltonian time evolution algorithms.

The quantum approximate optimization algorithm (QAOA) was first proposed in 2014 by Farhi, Goldstone and Gutmann to solve classical combinatorial optimization problems on near-term quantum devices \cite{qaoa}.
Since then, it has been investigated analytically \cite{qaoa_universality_0,qaoa_universality_1,bounded_depth_algs}, numerically \cite{qaoa_fermionic,variational_state_prep}, and on quantum computers \cite{qaoa_harvard_exp,qaoa_google_exp}.

In general, given a problem Hamiltonian $H_P$ that encodes the solution to the problem of interest onto $N$ qubits, QAOA applies two types of layers alternatingly to an initial state $\ket{+}^{\otimes N}$:
\begin{align}
    V_\text{QAOA}(\xx)=\prod_{j=p}^1 U_M(x_{2j})U_P(x_{2j-1}),
\end{align}
where $p$ is the number of blocks which determines the depth of the circuit, $U_M(x)=\exp\left(-i x H_M\right)$ with $H_M = \sum_{k=1}^N X_k$ is the so-called \emph{mixing layer}, and $U_P(x)=\exp(-ixH_P)$ is the time evolution under $H_P$. The parameters $\xx$ can then be optimized to try to minimize the objective function 
\begin{align}
    E(\xx) = \bra{+}^{\otimes N} V^\dagger_\text{QAOA}(\xx) H_P V_\text{QAOA}(\xx) \ket{+}^{\otimes N}.
\end{align}

Here we focus on the layer $U_P$, and we look at the example of $\maxcut$ in particular.
The corresponding problem Hamiltonian for an unweighted graph $G=(\mathcal{V}, \mathcal{E})$ with $N$ vertices $\mathcal{V}$ and $M$ edges $\mathcal{E}$ reads
\begin{align}
    H_P = \sum_{(a,b)\in \mathcal{E}} \frac{1}{2}(1-Z_aZ_b),
\end{align}
and $U_P$ correspondingly contains $M$ two-qubit Pauli-$Z$ rotations $R_{ZZ}$.

We note that $H_M$ has eigenvalues $-N, -N+2, \cdots, N$, which means the corresponding frequencies (differences of eigenvalues) are $2, \cdots, 2N$.
Thus, treating $U_M(x_{2j})$ as a single operation, Eq.~\eqref{eq:E_fourier2} implies that $E(\xx)$ can be considered an $N$-order trigonometric polynomial in $x_{2j}$, and the parameter-shift rules we derive in Sec.~\ref{sec:univariate_function_eq} will apply with $R=N$.
Similarly, $H_P$ has corresponding frequencies in the set $[M]$, and it will obey the parameter-shift rule for $R=M$, although we may be able to give better upper bounds $\lambda$ for $R$.
Thus the unique positive differences $\{\Omega_\ell\}$ for those layers, i.e., the frequencies of $E(\xx)$ with respect to the parameter $\{x_{2j-1}\}_{j\in[p]}$, take integer values within the interval $[0,\lambda]$ as well. We may therefore use Eq.~\eqref{eq:parshift_eq}, with the knowledge that $R\leq\lambda\leq M$.

Note that knowing \textit{all} frequencies of $E(x)$ requires knowledge of the full spectrum of $H_P$ --- and in particular of $\lambda$ --- which in turn is the solution of $\maxcut$ itself. As a consequence, the motivation for performing QAOA becomes obsolete.
Therefore, in general we cannot assume to know $\{\Omega_\ell\}$ (or even $R$), but instead require upper bounds $\varphi(G)\geq\maxcut(G)=\lambda$ which can be used to bound the largest frequency, and thus the number of frequencies $R$ and subsequently the number of terms in the parameter-shift rule.
It is noteworthy that even if the \emph{largest} frequency $\lambda$ is known exactly via a tight bound --- which restricts the Fourier spectrum to the integers $[\lambda]$ --- not \emph{all} integers smaller than $\lambda$ need to be present in the set of frequencies $\{\Omega_\ell\}$, so that the estimate for $R$ may be too large\footnote{A simple example for this is the case of $2k$-regular graphs; here, $H_P$ only has even eigenvalues, and therefore all frequencies are even as well. Given an upper bound $\varphi$, we thus know the number of frequencies to satisfy $R\leq\varphi/2$.}.

One way to obtain an upper bound uses analytic results based on the Laplacian of the graph of interest \cite{MaxCut_bounds, laplace_eigenvalues}, for which automatic bound generating programs exist \cite{automated_laplace_eigenvalues}.
An alternative approach uses semi-definite programs (SDPs) that solve relaxations of the $\maxcut$ problem, the most prominent being the \emph{Goemans-Williamson (GW)} algorithm \cite{GW_algorithm} and recent extensions thereof that provide tighter upper bounds \cite{MaxCut_SDP_geometry, MaxCut_SDP_tight}.
The largest eigenvalue is guaranteed to be within $\sim0.878$ of these SDP upper bounds.

To demonstrate the above strategy, we summarize the number of evaluations required for the gradient and Hessian of an $n$-parameter QAOA circuit on $N$ qubits for $\maxcut$ in Tab.~\ref{tab:qaoa_evaluations}, comparing the approach via decomposing the circuit, to the one detailed above based on $\varphi$ and the improved Hessian measurement scheme in Sec.~\ref{sec:cheaper_hessian}.
Here, we take into account that half of the layers are of the form $U_P$, and the other half are mixing layers with $R=N$ frequencies.
We systematically observe the number of evaluations for the gradient to be cut in half, and the those for the gradient and Hessian together to scale with halved order in $N$ (and $k$, for regular graphs).

\begin{table*}
    \centering
    \begin{tabular}{l|cc|ccc}
        \multirow{2}{*}{Graph type} & \multicolumn{2}{c|}{Decomposition-based} & \multicolumn{3}{c}{Gen. shift rule}\\
        \cline{2-6}
        & $\nnabla E$ & $\nnabla E \& \nnabla^{\otimes 2} E$ & Bound $\varphi$ & $\nnabla E$ & $\nnabla E \& \nnabla^{\otimes 2} E$  \\
        \hline
        General &  $(M+N)n$  & $\order{n^2(M+N)^2}$ & $\varphi$ & $n(\varphi+N)$ & $\order{n^2(\varphi+N)}$ \\
        Complete & $\frac{1}{2}n(N^2+N)$  & $\order{n^2N^4}$ & $\left\lfloor\frac{N^2}{4}\right\rfloor$ & $n\!\left(\left\lfloor\frac{N^2}{4}\right\rfloor+N\right)$ & $\order{n^2N^2}$ \\
        $2k$-regular & $(k+1)nN$ & $\order{k^2n^2N^2}$ & $kN$ & $\frac{k+2}{2}nN$ & $\order{kn^2N}$\\
        $(2k\!\!+\!\!1)$-regular & $\frac{2k+3}{2}nN$ & $\order{k^2n^2N^2}$ & $\frac{2k+1}{2}N$ & $\frac{2k+3}{2}nN$ & $\order{kn^2N}$\\
    \end{tabular}
    \caption{Evaluation numbers for the gradient, or both the gradient and the Hessian, for QAOA circuits for $\maxcut$ on several types of graphs.
    Each graph has $N$ vertices and a graph type-specific number $M$ of edges, and the (even) number of parameters is denoted as $n$.
    For $K$-regular graphs, we know $M=\min\{(N^2-N)/2, KN/2\}$, and the latter value is used in the displayed evaluation costs; if the former value forms the minimum, the graph is in fact complete.
    The left column is based on decomposing the circuit, applying the conventional two-term parameter-shift rule per elementary gate and iterating it for the Hessian.
    The right column employs the generalized parameter-shift rule Eq.~\eqref{eq:parshift_eq} combined with an upper bound $\varphi$ for the largest eigenvalue $\lambda$ of the problem Hamiltonian, as well as the reduced number of evaluations for Hessian off-diagonal terms from Sec.~\ref{sec:cheaper_hessian}. The bound $\varphi$ for complete graphs can be found in Ref.~\cite{MaxCut_bounds}.
    }
    \label{tab:qaoa_evaluations}
\end{table*}

\begin{figure}
    \centering
    \includegraphics[width=0.48\textwidth]{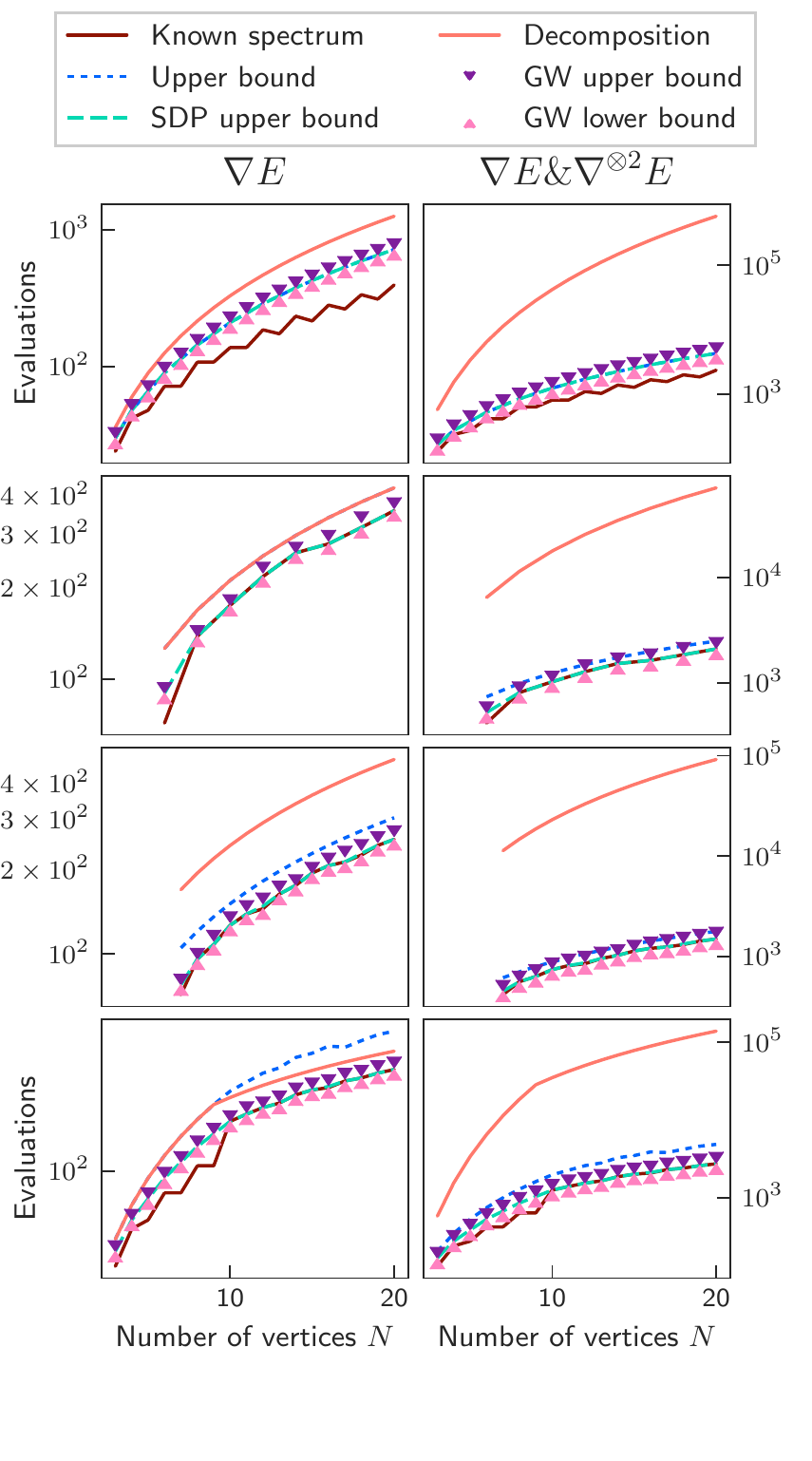}
    \caption{Evaluation numbers $N_\text{eval}$ for the gradient (\emph{left}) or both the gradient and the Hessian (\emph{right}) for $n=6$ parameter QAOA circuits for $\maxcut$ on graphs of several types and sizes.
    Using numerical upper bounds together with our new parameter-shift rule (GW -- \emph{purple triangles} and its generalization -- \emph{dashed turquoise}) reduces the resource requirements for both quantities significantly, compared to the previously available decomposition-based method (\emph{solid orange}).
    The rows correspond to the various considered graph types (\emph{top to bottom}): complete, $5$-regular, $6$-regular and (up to) $4N$ randomly sampled edges.
    The requirements for the decomposition-based approach and the analytic upper bound (\emph{dotted blue}) correspond to the results in the left and right column of Tab.~\ref{tab:qaoa_evaluations}, respectively.
    The numerical \emph{upper} bounds both use the minimized objective value of SDPs for relaxations of $\maxcut$ to obtain the bound $\varphi$, which depends on the graph instance.
    The GW-based \emph{lower} bound (\emph{pink triangles}) is obtained by randomly mapping the output state of the GW algorithm to $10$ valid cuts and choosing the one with the largest cut value.
    Note that $K$-regular graphs are only defined for $N>K$ and $NK \mod 2 =0$ and that graphs with $\kappa N$ sampled edges are complete for $N\leq 2\kappa +1$, leading to a change in the qualitative behaviour in the last row at $N=2\kappa+2=10$.
    }
    \label{fig:qaoa_evaluations}
\end{figure}

In addition, we display the numbers of circuit evaluations from Tab.~\ref{tab:qaoa_evaluations} together with SDP-based bounds for $\lambda$ and the true minimal number of evaluations required for the parameter-shift rule in Fig.~\ref{fig:qaoa_evaluations}.
For this, we sampled random unweighted graphs of the corresponding type and size and ran the GW algorithm as well as an improvement thereof to obtain tighter bounds \cite{MaxCut_SDP_geometry}.
On one hand we observe the advantage of the generalized parameter-shift rule and the cheaper Hessian method that can be read off already from the scalings in Tab.~\ref{tab:qaoa_evaluations}.
On the other hand, we find both SDP-based upper bounds to provide an exact estimate of the largest eigenvalue in the $N \le 20$ regime, as can be seen from the cut values obtained from the GW algorithm that coincide with the upper bound.
In cases in which the frequencies $\{\Omega_\ell\}$ occupy all integers in $[R]$, this leads to an exact estimate of $R$ and the evaluations in the shift rule.
For all graph types but complete graphs, the SDP-based upper bounds yield a better estimate for the number of terms than the respective analytic bound $\varphi$, which improves the generalized shift rule further.

In summary, we find the generalized parameter-shift rule to offer a constant prefactor improvement when computing the gradient and an improvement of at least $\order{N}$ when computing both the gradient and the Hessian. 
For certain graph types, knowledge about the structure of the spectrum and tight analytic bounds provide this advantage already, whereas for other graph types the SDP-based bounds reduce the evaluation numbers significantly.

\subsection{Rotosolve}\label{sec:rotosolve}

The \emph{Rotosolve} algorithm is a coordinate descent algorithm for minimizing quantum cost functions. It has been independently discovered multiple times \cite{pqc_calculus,sequential_minimal_optimization,jacobi_and_anderson,rotosolve}, with \cite{rotosolve} giving the algorithm its name but only (along with \cite{jacobi_and_anderson}) considering parametrized Pauli rotations, and \cite{pqc_calculus,sequential_minimal_optimization} covering other unitaries with integer-valued generator eigenvalues.

The Rotosolve algorithm optimizes the rotation angles sequentially: for one variational parameter $x_k$ at a time, the cost function is reconstructed as a function of that parameter using $2R_k+1$ evaluations, the minimum of the reconstruction is calculated, and then the parameter is updated to the minimizing angle. For the case of Pauli rotation gates this minimum can be found via a closed-form expression. Recent studies have shown such coordinate descent methods to work well on many tasks \cite{layerwise_QNN_training,rotosolve,sequential_minimal_optimization,hardware_VQA_TE}, although there are limited cases where these methods fail \cite{training_transitions}.

While Rotosolve is not gradient-based, our cost reduction for the gradient presented in Sec.~\ref{sec:qaoa_truncation}
stems from a cost reduction for function reconstruction, and hence is applicable to Rotosolve as well.

As shown in Sec.~\ref{sec:full_reconstruction}, the univariate objective function can also be fully reconstructed if the parametrized unitaries are more complicated than Pauli rotations, using the function value itself and the evaluations from the generalized parameter-shift rule.
Since the generalized parameter-shift rule also applies for non-equidistant frequencies (see App.~\ref{sec:gen}), the reconstruction works in the same way for arbitrary single-parameter gates.
This extends our generalization of Rotosolve beyond the previously known integer-frequency case \cite{pqc_calculus,sequential_minimal_optimization}, although the number of frequencies---and thus the cost---for the reconstruction are typically significantly increased for non-integer frequencies.
While the minimizing angle might not be straightforward to express in a closed form as it is the case for a single frequency, the one-dimensional minimization can efficiently be carried out numerically to high precision, via grid search or semi-definite programming \cite[Chapter~4.2]{convex_optimization}.

\subsection{Quantum analytic descent}\label{sec:qad}
Quantum analytic descent (QAD) \cite{qad} also approaches the optimization problem in VQAs via trigonometric interpolation.
In contrast to Rotosolve, it considers a model of all parameters \textit{simultaneously} and
includes second-order derivatives, but this model only is a \emph{local approximation} of the
full cost function.
Additionally, QAD has been developed for circuits that exclusively contain Pauli rotations as parametrized gates.

The algorithm evaluates the cost function $E$ at $2n^2+n+1$ points around a reference point $\xx_0$, and then constructs a trigonometric model of the form\footnote{We slightly modify the trigonometric basis functions from Ref.~\cite{qad} to have leading order coefficients $1$.}
\begin{align}
    \hat{E}(\xx_0+\xx)&=A(\xx)\left[E^{(A)}+2\EE^{(B)}\cdot \tan\left(\frac{\xx}{2}\right)\right.\nonumber\\
    &+2\EE^{(C)}\cdot \tan\left(\frac{\xx}{2}\right)^{\odot 2}\\
    &\left.+4\tan\left(\frac{\xx}{2}\right)\cdot E^{(D)}\cdot \tan\left(\frac{\xx}{2}\right)\right],\nonumber
\end{align}
Here, we introduced $A(\xx)\coloneqq\prod_k \cos^2\left(\frac{x_k}{2}\right)$ and the element-wise square of a vector $\vv$, $(\vv^{\odot 2})_k\coloneqq v_k^2$ as for the Hessian diagonal.
The coefficients $E^{(A/B/C/D)}$ are derived from the circuit evaluations, taking the form of a scalar, two vectors and an upper triangular matrix.
More precisely, the expansion basis is chosen such that $\EE^{(B)}=\nnabla E(\xx_0)$, $\EE^{(C)}=\nnabla^{\odot 2} E(\xx_0)$, and $E^{(D)}$ is the strictly upper triangular part of the Hessian.
Note that for this model $2n^2+n+1$ evaluations are used to obtain $n^2/2+3n/2+1$ parameters.
In the presence of statistical noise from these evaluations, it turns out that building the model to a desired precision and inferring modelled gradients close to the reference point $\xx_0$ has resource requirements similar to measuring the gradient directly \cite{qad}.

This model coincides with $E(\xx)$ at $\xx_0$ up to second order, and in the vicinity its error scales with the third order of the largest parameter deviation \cite{qad}.
After the construction phase, the model cost is minimized in an inner optimization loop, which only requires classical operations.
For an implementation and demonstration of the optimization, we also refer the reader to \cite{qad_demo} and \cite{qad_repo}.

In the light of the parameter-shift rules and reconstruction methods, we propose three (alternative) modifications of QAD.
The first change is to reduce the required number of evaluations.
As the coefficients $E^{(A/B/C/D)}$ consist of the gradient and Hessian, they allow us to exploit the reduced resource requirements presented in Tab.~\ref{tab:quantity_cost}
\footnote{In addition, we may skip the $n$ evaluations with shift angle $\pi$ proposed in Ref.~\cite{qad}, and instead measure the Hessian diagonal as discussed in Sec.~\ref{sec:cheaper_hessian}.}.
In the case originally considered by the authors, i.e., for Pauli rotations only, this reduces the number of evaluations from $2n^2+n+1$ to $(3n^2+n)/2+1$.

A second, alternative modification of QAD is to keep all evaluations as originally proposed to obtain the full second-order terms, i.e., we may combine the shift angles for each pair of parameters, and use them for coefficients of additional higher-order terms.
This extended model (see App.~\ref{sec:derivation_extQAD}) has the form
\begin{align}\label{eq:extQAD}
    \mathring{E}(\xx_0+\xx)&=\hat{E}(\xx_0+\xx)+4A(\xx)\tan\left(\frac{\xx}{2}\right)^{\odot 2}\\
    &\cdot\left[E^{(F)}\cdot \tan\left(\frac{\xx}{2}\right) +E^{(G)}\cdot\tan\left(\frac{\xx}{2}\right)^{\odot 2} \right],\nonumber
\end{align}
where $E^{(F)}$ is symmetric with zeros on its diagonal and $E^{(G)}$ is a strictly upper triangular matrix.
This extended model has $2n^2+1$ degrees of freedom, which matches the number of evaluations exactly.

While the QAD model reconstructs the univariate restrictions of $E$ to the coordinate axes correctly, the extended model $\mathring{E}$ does so for the bivariate restrictions to the plane spanned by any pair of coordinate axes.
It remains to investigate whether and for which applications the extension yields a better optimization behaviour;
for functions in which pairs of parameters yield a good local approximation of the landscape, it might provide an improvement.

The third modification we consider is to generalize the previous, extended QAD model to \textit{general} single-parameter quantum gates.
This can be done via a full trigonometric interpolation to second order, which is detailed in App.~\ref{sec:derivation_genQAD}, exactly reconstructing the energy function when restricted to any coordinate plane at the price of $2(\norm{\RR}_1^2-\norm{\RR}_2^2+\norm{\RR}_1)+1$ evaluations.

\begin{figure*}
    \centering
    \includegraphics[width=\textwidth]{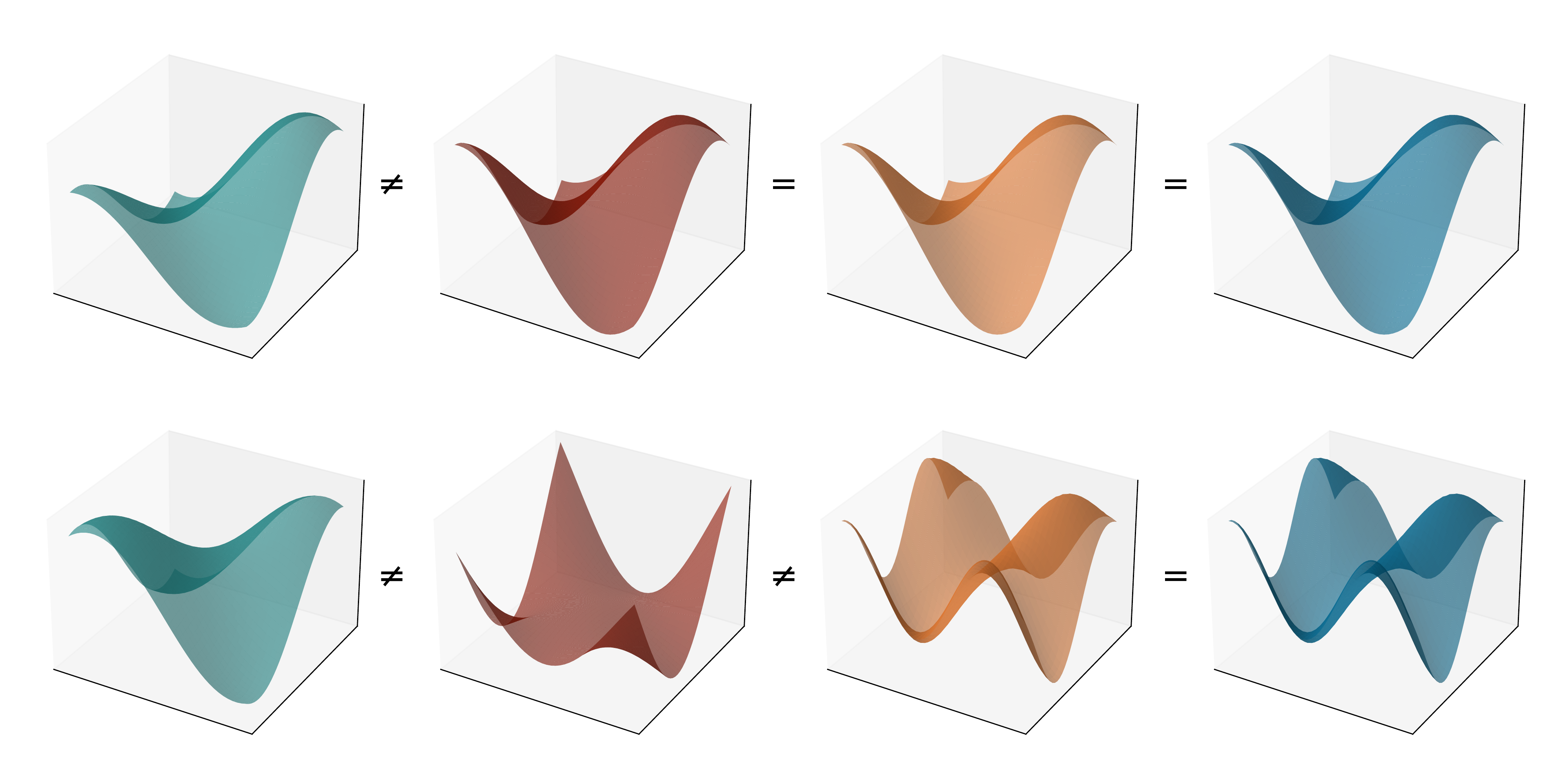}
    \caption{The QAD model (\emph{left}), its extension $\mathring{E}$, see Eq.~\eqref{eq:extQAD}, that includes full second-order terms (\emph{center left}), and the second-order trigonometric interpolation model (\emph{center right}), as well as the original expectation value $E$ (\emph{right}).
    The original function is generated from toy Hamiltonians in a two-parameter example circuit, with one frequency (\emph{top}) and two frequencies (\emph{bottom}) per parameter.
    The QAD model produces a local approximation to $E$ that deviates away from $\xx_0$ at a slow rate for $R=1$ but faster for $R=2$.
    The extension $\mathring{E}$ reuses evaluations made for the Hessian to capture the full bivariate dependence for a single frequency but is not apt to model multiple frequencies either.
    Finally, the trigonometric interpolation generalizes $\mathring{E}$. This means it coincides with $\mathring{E}$ for $R=1$, but also reproduces the full bivariate function for $R>1$.
    }
    \label{fig:qad_landscapes}
\end{figure*}

Using toy model circuits and Hamiltonians, we demonstrate the qualitative difference between the QAD model, its extension $\mathring{E}$, and the generalization to multiple frequencies in Fig.~\ref{fig:qad_landscapes}.

\section{Discussion} \label{sec:discussion}
In this work, we derive interpolation rules to exactly express quantum functions $E(x)$ as a linear combination of evaluations $E(x_\mu)$, assuming $E(x)$ derives from parametrized gates of the form $U(x) = \exp(ixG)$. Our method relies on the observation that $E(x)$ can be expressed as trigonometric polynomial in $x$, characterized by a set of $R$ \emph{frequencies} that correspond to distinct differences in the eigenvalues of $G$. This observation allows us to derive our results using trigonometric interpolation methods.

In addition to a full reconstruction of $E(x)$, the presented approach offers parameter-shift rules for derivatives of arbitrary order and recipes to evaluate multivariate derivatives more cheaply.
Using the concept of the stochastic parameter-shift rule, quantum gates of the form $U_F(x)=\exp(i(xG+F))$ can be differentiated as well.

Nevertheless, much remains unknown about the practicality of our new parameter-shift rules.
For the common case that we have $R$ equidistant frequencies, Sec.~\ref{sec:comparison_eq} shows that the scaling of the required resources is similar between na\"ively applying our generalized parameter-shift rules, and applying parameter-shift rules to a decomposition of $U(x)$.
This holds for the first derivative and also for the required shot budget when computing the second derivative, whereas the number of unique circuits is significantly smaller for the new, generalized shift rule.

Our observations lead to several open questions:
\begin{itemize}[leftmargin=10pt, label={\raisebox{0.25ex}{\tiny$\bullet$}}]
    \item In which situations can we obtain better bounds on the number of frequencies? We investigated an example for QAOA in Sec.~\ref{sec:qaoa_truncation}, but are there other examples?
    \item For general $G$ (e.g., $G = \sum_j c_j P_j$ with real $c_j$ and Pauli words $P_j$), the frequencies will not be equidistant, and in fact $R$ may scale quadratically in the size of $U$. Na\"ively applied, our method would then scale poorly compared to decomposing $G$. Can we apply an approximate or stochastic parameter-shift rule with a better scaling?
    \item Would it ever make sense to \emph{truncate} these parameter-shift rules to keep only terms corresponding to smaller frequencies? This is inspired by the idea of using low-pass filters to smooth out rapid changes of a signal.
    \item Our work on function reconstruction extends QAD to all gates with equidistant frequencies. Similarly, it allows Rotosolve, which has been shown to work remarkably well on some applications, to be used on all quantum gates with arbitrary frequencies.
    Is there a classification of problems on which these model-based algorithms work well?
    And can we reduce the optimization cost based on the above ideas?
    \item More generally, can we apply the machinery of Fourier analysis more broadly, e.g., to improve optimization methods in the presence of noise?
\end{itemize}
We hope that this work serves as an impetus for future work that will further apply signal processing methods to the burgeoning field of variational quantum computing.

\section*{Acknowledgements}
We would like to thank Nathan Killoran, Maria Schuld, Matthew Beach, and Eric Kessler for helpful comments on the manuscript, as well as Christian Gogolin and Gian-Luca Anselmetti for valuable discussions.

\section*{Code availability}
The scripts used to create the data and plots for Figs.~\ref{fig:qaoa_evaluations} and~\ref{fig:qad_landscapes} can be found at \cite{repo}.

\bibliographystyle{bibstyle2.bst}
\bibliography{main}


\begin{appendix}
\section{Technical derivations}\label{sec:derivations}
\subsection{Derivation of explicit parameter-shift rules}\label{sec:derivation_dirichlet}
Here we derive the trigonometric interpolation via Dirichlet kernels.

\subsubsection{Full reconstruction}\label{sec:derivation_full_reconstruction}
We start out by exactly determining $E(x)$ given its value at points $\{x_\mu = \frac{2\mu}{2R+1}\pi\}, \mu \in \{-R, \cdots, R\}$. This is a well-known problem \cite[Chapter~X]{trig_interpolation}; we reproduce the result below for completeness.

Consider the \emph{Dirichlet kernel}
\begin{align}
\D(x) &= \frac{1}{2R+1} + \frac{2}{2R+1} \sum_{\ell=1}^{R}\cos(\ell x) \\
&= \frac{\sin\left(\frac{2R+1}{2}x\right)}{(2R+1)\sin \left(\frac{1}{2} x\right)}
\end{align}
where the limit $x\rightarrow 0$ is taken when evaluating $\D(0)$. The functions $\D(x-x_\mu)$ are linear combinations of the basis functions $\{\sin(\ell x)\}_{\ell \in [R]}$, $\{\cos(\ell x)\}_{\ell \in [R]_0}$, and they satisfy $\D(x_{\mu'} - x_\mu) = \delta_{\mu\mu'}$. Therefore it is evident that 
\begin{align}
E(x) &= \sum_{\mu=-R}^R E(x_\mu) \D(x-x_\mu) \\
&= \frac{\sin\left(\frac{2R+1}{2}x\right)}{2R+1} \sum_{\mu=-R}^R E\left(x_\mu\right) \frac{(-1)^\mu} {\sin \left(\frac{x-x_\mu}{2}\right)}. \label{eq:full_reconstruction}
\end{align}

As an example, for $R=1$ (e.g., when the generator $G$ satisfies $G^2 = \id$) we have the formula
\begin{align}
    E(x) = \frac{\sin\left(\frac{3}{2}x\right)}{3} &\left[- \frac{E(-\frac{2}{3}\pi)}{\sin(\frac{x}{2}+\frac{\pi}{3})}\right.\\
    &\left.+\frac{E(0)}{\sin(\frac{x}{2})} - \frac{E(\frac{2}{3}\pi)}{\sin(\frac{x}{2}-\frac{\pi}{3})}  \right].\nonumber
\end{align}
Derivatives of $E(x)$ can be straightforwardly extracted from this full reconstruction.

\subsubsection{Odd kernels}
We now consider the case of determining
$E_\text{odd}$ given its value at evenly spaced points $\{x_\mu = \frac{2\mu-1}{2R}\pi\}_{\mu\in[R]}$ \footnote{Unlike Sec.~\ref{sec:derivation_full_reconstruction}, we are not aware of a prior reference for the derivations for this subsection (reconstructing the odd part) and the next (reconstructing the even part).}.
Consider the \emph{modified Dirichlet kernel}:
\begin{align}
\Dmod(x) &= \frac{1}{2R} + \frac{1}{2R}\cos(Rx) + \frac{1}{R} \sum_{\ell=1}^{R-1}\cos(\ell x) \\
&= \frac{\sin(Rx)}{2R\tan \left(\frac{1}{2} x\right)}
\end{align}
where we again assume the limit $x\rightarrow 0$ is taken when evaluating $\Dmod(0)$.
This kernel satisfies the relations
\begin{equation}
\Dmod(x_{\mu'} - x_\mu) = \delta_{\mu\mu'}, \quad \Dmod(x_{\mu'}+x_\mu)=0,
\end{equation}
but unfortunately, $\Dmod(x)$ is a linear combination of cosines, not sines; it's an even function, not an odd function.
We therefore instead consider the linear combinations
\begin{align}
\Dodd_\mu(x) &\coloneqq \Dmod(x-x_\mu) - \Dmod(x+x_\mu) \\
&= \frac{\sin(R (x-x_\mu))}{2R \tan\left(\frac{1}{2} (x-x_\mu)\right)} - \frac{\sin(R (x+x_\mu))}{2R \tan\left(\frac{1}{2} (x+x_\mu)\right)} \nonumber\\
&= \frac{1}{R} \cos(x_\mu) \left[\frac{1}{2}\sin(Rx) + \sum_{\ell=1}^{R-1} \sin(\ell x) \right].\nonumber
\end{align}
Similarly to $\Dmod$, this kernel satisfies $\Dodd_\mu(x_{\mu'}) = \delta_{\mu\mu'}$ but it's a linear combination of the odd basis functions $\sin(\ell x),\ell\in[R]$.
Following from these two properties, we know that
\begin{align}
E_\text{odd}(x) &= \sum_{\mu=1}^R E_\text{odd}(x_\mu) \Dodd_\mu(x) \\
&= \sum_{\mu=1}^R \frac{E_\text{odd}(x_\mu)}{2R} \nonumber\\
&\quad\times\left[\frac{\sin(R (x-x_\mu))}{\tan\left(\frac{1}{2} (x-x_\mu)\right)}-\frac{\sin(R (x+x_\mu))}{\tan\left(\frac{1}{2} (x+x_\mu)\right)}\right]\nonumber
\end{align}
and we thus can reconstruct $E_\text{odd}$ with the $R$ evaluations $E_\text{odd}(x_\mu)$.

We also can extract from here a closed-form formula for the derivative at $x=0$, as it only depends on the odd part of $E$. We arrive at the \emph{general parameter-shift rule}:
\begin{align}
E'(0) &=\sum_{\mu=1}^R E_\text{odd}(x_\mu)\Dodd_{\mu}'(0)\\
&= \sum_{\mu=1}^R E_\text{odd}(x_\mu) \frac{\sin(Rx_\mu)}{2R\sin^2(\frac{1}{2}x_\mu)} \\
 &= \sum_{\mu=1}^R E_\text{odd}\left(\frac{2\mu-1}{2R}\pi\right) \frac{(-1)^{\mu-1}}{2R\sin^2\left(\frac{2\mu-1}{4R}\pi\right)}.\nonumber
\end{align}

Similarly, as the higher-order derivatives of $\Dodd_\mu$ can be computed analytically, we may obtain derivatives of $E$ of higher odd orders.

\subsubsection{Even kernels}
Next we reconstruct the even part $E_\text{even}$ again using the kernel $\Dmod(x)$ from above but choosing the $R+1$ points $x_\mu=\mu\pi/R$ for $\mu\in[R]_0$.
As the spacing between these points is the same as between the previous $\{x_\mu\}$, we again have $\Dmod(x_{\mu'}-x_\mu)=\delta_{\mu\mu'}$; but note we cannot directly use $\Dmod(x-x_\mu)$ as our kernel because $\Dmod(x-x_\mu)$ is an even function in $x-x_\mu$ but not in $x$. Instead we take the even linear combination
\begin{align}
\Deven_\mu(x) \coloneqq 
\begin{cases}
   \Dmod(x) & \text{if } \mu = 0\\
   \Dmod(x-x_\mu) + \Dmod(x+x_\mu) & \text{if } 0<\mu<R\\
   \Dmod(x-\pi) & \text{if } \mu = R\ .
\end{cases}\nonumber
\end{align}
Then the $\Deven_\mu$ are even functions and satisfy $\Deven_\mu(x_{\mu'}) = \delta_{\mu\mu'}$, leading to
\begin{align}
E_\text{even}(x) &= \sum_{\mu=0}^R E_\text{even}(x_\mu) \Deven_\mu(x).
\end{align}

The second derivative of $\Dmod$ is
\begin{align}
    {\Dmod}''(x) &= \frac{\sin(Rx)\left[1-2R^2\sin^2(\frac{1}{2}x)\right]}{4R\tan(\frac{1}{2}x)\sin^2(\frac{1}{2}x)} -\frac{\cos(Rx)}{2\sin^2 \left(\frac{1}{2}x\right)}\nonumber
\end{align}
and if we take the limit $x \rightarrow 0$:
\begin{align}
    {\Dmod}''(0) = -\frac{2R^2+1}{6}.
\end{align}
This yields the explicit parameter-shift rule for the second derivative:
\begin{align}
    E''(0) &= -E_\text{even}(0)\frac{2R^2+1}{6} + E_\text{even}(\pi) \frac{(-1)^{R-1}}{2} \nonumber\\
    &\quad +\sum_{\mu=1}^{R-1} E_\text{even}\left(\frac{\mu\pi}{R}\right)\frac{(-1)^{\mu-1}}{\sin^2 \left(\frac{\mu\pi}{2R}\right)}.
\end{align}
Again, derivatives of $E$ of higher even order can be computed in a similar manner, using the same evaluations $E_\text{even}\left(\frac{\mu\pi}{R}\right)$.

\subsection{Hessian parameter-shift rule}\label{sec:derivation_hessian_rule}
Here we consider the spectrum of the function
\begin{align}
    E^{(km)}(x) \coloneqq E(\xx_0+x\vv_{k,m}),
\end{align}
with $\vv_{k,m}=\vv_k+\vv_m$.
Without loss of generality, we assume $U_k$ to act first within the circuit and set $\xx_0=\boldsymbol{0}$.
As for the univariate case in Sec.~\ref{sec:derivation_funcform}, we may explicitly write the cost function as
\begin{align}
    E^{(km)}(x)&=\bra{\psi}U_k^\dagger(x)V^\dagger U_m^\dagger(x) B U_m(x)V U_k(x)\ket{\psi}\nonumber\\
    &=\sum_{j_1,\dots j_4=1}^d \overline{\psi_{j_1} v_{j_2j_1}} b_{j_2j_3} v_{j_3j_4} \psi_{j_4}\\
    &\times \exp\left({i\left(\omega_{j_4}^{(k)}-\omega_{j_1}^{(k)}+\omega_{j_3}^{(m)}-\omega_{j_2}^{(m)}\right)x}\right)\nonumber,
\end{align}
where $\omega^{(k,m)}$ are the eigenvalues of the generators of $U_k$ and $U_m$, respectively, and we denoted the entries of matrices by lowercase letters as before.
We may read off the occuring frequencies in this Fourier series in terms of the unique positive differences $\Omega^{(k,m)}$, leading to $\delta\Omega_{l_1l_2}=\pm\Omega_{l_1}^{(k)}\pm\Omega_{l_2}^{(m)}$.
We again only collect the positive values as they come in pairs\footnote{That is, for any $\delta\Omega$, we also have $-\delta\Omega$ in the Fourier series, and the representation as real-valued function subsums the two frequencies.}.

In case of integer-valued frequencies, there are $R_{km}=R_k+R_m$ such positive frequencies, namely all integers in $[R_k+R_m]$.
For arbitrary frequencies, all $\{\delta\Omega\}$ might be unique and we obtain up to $R_{km}=2R_kR_m+R_k+R_m$ frequencies.
Rescaling the smallest frequency enforces a small degree of redundancy so that $R_{km}=2R_kR_m+R_k+R_m-2$ is always achievable; for some scenarios specific rescaling factors might drastically reduce $R_{km}$ \footnote{Recall that we used rescaling for the equidistant frequency case to arrive at integer-valued $\{\Omega\}$, which in turn made the significant reduction above possible.}.

\subsection{Hadamard tests for the metric tensor}\label{sec:derivation_Hadamard_test}
In order to compute the metric tensor as the Hessian of the overlap $f(\xx)=-\frac{1}{2}|\!\braket{\psi(\xx)|\psi(\xx_0)}\!|^2$, we need to evaluate it at shifted positions $\xx= \xx_0+x \vv_{k,m}$.
This can be done by executing the circuit $V(\xx_0)$ and the adjoint circuit $V^\dagger(\xx)$ at the shifted position, and returning the probability to measure the $\mathbf{0}$ bitstring in the computational basis.
As all operations after the latter of the two parametrized gates of interest cancel between the two circuits, those operations can be spared, but the maximal depth is (almost) the doubled depth of $V$.

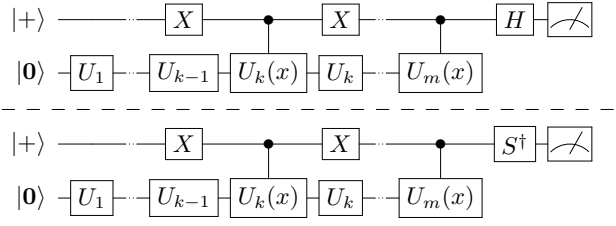
\begin{figure}
\flushright
\resizebox{\columnwidth}{!}{
    \Qcircuit @C=5pt @R=.7em {
    &\lstick{\ket{+}}            & \qw        & \qw & \qwdash & \gate{X}       & \ctrl{1}      &
    \gate{X}   & \qw & \qwdash & \ctrl{1}        & \gate{H} & \meter{Z}\\
    &\lstick{\ket{\boldsymbol{0}}}& \gate{U_1} & \qw & \qwdash & \gate{U_{k-1}} & \gate{U_k(x)} &
    \gate{U_k} & \qw & \qwdash & \gate{U_{m}(x)} &\\
    \ar @{--} []+<-18pt,0em>;[]+<\columnwidth,0em>&\\
    &\lstick{\ket{+}}            & \qw        & \qw & \qwdash & \gate{X}       & \ctrl{1}      &
    \gate{X}   & \qw & \qwdash & \ctrl{1}        & \gate{S^\dagger} & \meter{Z}\\
    &\lstick{\ket{\boldsymbol{0}}}& \gate{U_1} & \qw & \qwdash & \gate{U_{k-1}} & \gate{U_k(x)} &
    \gate{U_k} & \qw & \qwdash & \gate{U_{m}(x)} &
    }
}
    \caption{Circuits for the Hadamard tests to measure the overlap in Eq.~\eqref{eq:hadamard_test_overlap}, adapted from \cite[Fig.~5]{variational_ITE}.
    The basis rotation in the last operation on the auxiliary qubit determines whether the real (\emph{top}) or the imaginary (\emph{bottom}) part of $\braket{\psi(\xx_0+x\vv_{k,m})|\psi(\xx_0)}$ is calculated.
    All unitaries without argument are understood as $U_j=U_j((\xx_0)_j)$.
    }
    \label{fig:hadamard_test_parshift}
\end{figure}

Alternatively, we may use a Hadamard test as derived in the appendix of Ref.~\cite{variational_ITE}.
There, it was designed to realize the derivative overlaps $\real{\braket{\partial_k\psi(\xx)|\partial_m\psi(\xx)}}$ for the metric tensor directly, assuming the generator to be a Pauli word and therefore unitary.
However, it can also be used to calculate the real or imaginary part of 
\begin{align}\label{eq:hadamard_test_overlap}
    \braket{\psi(\xx)|\psi(\xx_0)} &= \bra{\boldsymbol{0}} U_1^\dagger((\xx_0)_1)\cdots U_k^\dagger((\xx_0)_k+x) \nonumber\\
    &\hspace{-0.5cm}\cdots U_{m-1}^\dagger((\xx_0)_{m-1}) U_m^\dagger(x) U_{m-1}((\xx_0)_{m-1})\nonumber\\
    &\hspace{-0.5cm}\cdots U_1((\xx_0)_1)\ket{\boldsymbol{0}}.
\end{align}
by measuring the auxiliary qubit in the $Z$ or $Y$ basis.
The corresponding circuit is shown in Fig.~\ref{fig:hadamard_test_parshift}.

While the original proposal has to split up the generators into Pauli words and implement one circuit per combination of Pauli words from $x_k$ and $x_m$, the number of circuits here is dictated by the number of evaluations in the parameter-shift rule.
In order to measure $f(\xx)$, the real and the imaginary part both have to be measured, doubling the number of circuits.

\subsection{Coefficient norms for univariate derivatives via equidistant shifts}\label{sec:derivation_coefficient_norms}
The $\ell_1$-norm of the coefficients in parameter-shift rules dictates the number of shots required to reach certain precision (see Sec.~\ref{sec:comparison_discussion}). Here, we explicitly compute this norm for both the general and decomposition-based parameter-shift rule for the first- and second-order univariate derivative.
For the entire analysis, we approximate the single-shot variance $\sigma^2$ to be constant as detailed in the main text. 

\subsubsection{Norm for general parameter-shift rule}\label{sec:derivation_parshift_coefficient_norm}
For the case of equidistant shift angles, we can compute the norm of the coefficient vector $\yy^{(1,2)}$ in the parameter-shift rules in Eqs.~(\ref{eq:parshift_eq},\ref{eq:parshift_eq2}) explicitly, in order to estimate the required shot budget for the obtained derivative.
For the first order, we note that the evaluations of $E$ come in pairs, with the same coefficient up to a relative sign.
This yields (recalling that $x_{\mu}=\frac{2\mu-1}{4R}\pi$):
\begin{align}\label{eq:parshift_coefficient_norm_1}
    \norm{\yy^{(1)}}_1 &= \frac{1}{2R}\sum_{\mu=1}^R \frac{1}{\sin^2(x_\mu)} = R,
\end{align}
which follows from $\sin^{-2}(x_\mu)=\cot^2(x_\mu)+1$ and \cite[Formula~(445)]{summation_of_series}:
\begin{align}
    \sum_{\mu=1}^R \cot^2(x_\mu) &=2R^2-R. \label{eq:cot_sum}
\end{align}
A derivation for Eq.~\eqref{eq:cot_sum} can be adapted from Ref.~\cite{SE_sum_of_inv_tan_squared}, which we present below for completeness:
\begin{align*}
    -i (-1)^{\mu} &= \exp(i 2Rx_\mu)\\
              &= \Big(\cos(x_\mu)+i\sin(x_\mu)\Big)^{2R}\\
              &= \sum_{r=0}^{2R}\binom{2R}{r}\left(\cos(x_\mu)\right)^{2R-r}\left(i\sin(x_\mu)\right)^{r}\\
\Rightarrow \quad 0 &= \sum_{r=0}^{R}\binom{2R}{2r}\left(\cos(x_\mu)\right)^{2R-2r}\left(i\sin(x_\mu)\right)^{2r}\\
              &= \sum_{r=0}^{R}\binom{2R}{2r}\Big(-\cot^2(x_\mu)\Big)^{R-r}
\end{align*}
Here we have applied the binomial theorem, extracted the real part, and divided by $(i\sin(x_\mu))^{2R}$ (note that $0<x_\mu<\pi/2$).
From the last equation above, we see that $\cot^2(x_\mu)$ is a root of the function $g(\chi)=\sum_{r=0}^R\binom{2R}{2r}(-\chi)^{R-r}$ for all $\mu\in[R]$.
As $g$ is a polynomial of degree $R$, we thus know \emph{all} its roots and may use the simplest of Vieta's formulas:
\begin{align}
    \sum_{\mu=1}^R \tau_{\mu} &= -\frac{g_{R-1}}{g_R}
\end{align}
with roots $\{\tau_{\mu}\}_\mu$ of $g$, and $g_{j}$ the $j$th order Taylor coefficient of $g$.
Plugging in the known roots and coefficients we get
\begin{align}
    \sum_{\mu=1}^R \cot^2(x_\mu) &= -\frac{(-1)^{R-1}\binom{2R}{2}}{(-1)^{R}\binom{2R}{0}}\\
                                  &=2R^2-R.
\end{align}

For the second order we may repeat the above computation with small modifications\footnote{Recall that the angles differ between the two derivatives.}, arriving at 
$g(\chi) = \sum_{r=0}^{R-1}\binom{2R}{2r+1}(-\chi)^{R-r}$
and therefore at 
\begin{align}
    \norm{\yy^{(2)}_1} &= \frac{2R^2+1}{6}+\frac{1}{2}+(R-1)-\frac{(-1)^{R-1}\binom{2R}{3}}{(-1)^R\binom{2R}{1}}\nonumber\\
    &=R^2.
\end{align}

\subsubsection{Norm for decomposition}
\label{sec:derivation_decomp_coefficient_norm}
If we compute the first- and second-order derivatives via a decomposition that contains $\p$ parametrized elementary gates, we need to apply the original two-term parameter-shift rule to each of these gates separately.
For the first-order derivative, we simply sum all elementary derivatives.
For integer-valued frequencies, $x$ typically feeds without prefactor into the gates in the decomposition, so that the decomposition-based shift rule reads
\begin{align}
    E'(0)=\frac{1}{2\sin(x_1)}\sum_{k=1}^\p [E^{(k)}(x_1)-E^{(k)}(-x_1)],
\end{align}
where $E^{(k)}$ denotes the cost function based on the decomposition, in which only the parameter of the $k$th elementary gate is set to the shifted angle $x_1$ and to $0$ in all other gates.
To maximize $\sin(x_1)$, we choose $x_1=\pi/2$, and as a reuslt all $2\p$ coefficients have magnitude $1/2$, and therefore
\begin{align}
    \norm{\yy_\text{decomp}^{(1)}}_1=\p.
\end{align}
Due to all coefficients being equal, the optimal shot allocation is $N/(2\p)$ for all terms.

For the second-order derivative, the full Hessian has to be computed from the decomposition as described in Ref.~\cite{Higher_order_derivatives} and all elements have to be summed\footnote{Here we do not anticipate the cheaper Hessian evaluation from Sec.~\ref{sec:cheaper_hessian}.}:
\begin{align}
    E''(0)&=\frac{1}{2\sin^2(x_1)}\sum_{\substack{k,m=1\\k< m}}^\p\\
    &\bigg[E^{(km)}(x_1,x_1)-E^{(km)}(-x_1,x_1)\nonumber\\
    &-E^{(km)}(x_1,-x_1)+E^{(km)}(-x_1,-x_1)\bigg] \nonumber\\
    &+\frac{1}{2}\sum_{k=1}^\p [E^{(k)}(\pi)-E(0)]\nonumber
\end{align}
where $E^{(km)}(x_1, x_2)$ is defined analogously to $E^{(k)}$ but the shift angles put into the $k$th and $m$th elementary gate may differ.
Fixing the shift angle to $\pi/2$ again, we have $2 \p(\p-1)$ coefficients of magnitude $1/2$ for the off-diagonal terms, $\p$ coefficients of magnitude $1/2$ for the $E^{(k)}(\pi)$ and one coefficient with magnitude $\p/2$ for $E(0)$, summing to
\begin{align}
    \norm{\yy_\text{decomp}^{(1)}}_1=2\p(\p-1)\frac{1}{2}+\p\frac{1}{2}+\frac{\p}{2}=\p^2.
\end{align}
Here the optimal shot allocation is to measure all shifted terms with $N/(2\p^2)$ shots, and $E(0)$ with $N/(2\p)$ shots.

\subsection{Coefficient norms for the Hessian}
\label{sec:derivation_coefficient_norms_hessian}
Similar to the previous section, we compute the coefficient norms for three methods to compute the Hessian for equidistant frequencies and shifts:
We may use the diagonal shift rule in Eq.~\eqref{eq:hessian_parshift}, repeat the general parameter-shift rule, or decompose the circuit and repeat the original parameter-shift rule.
For the first approach, the diagonal entries of the Hessian---and thus the shifted evaluations for those entries---are reused to compute the off-diagonal ones, whereas the shifted evaluations for the repeated shift rule are distinct for all Hessian entries.
This difference makes the cost comparison for a single Hessian entry difficult.
We therefore consider the root mean square of the Frobenius norm of the difference between the true and the estimated Hessian as quality measure.
The matrix of expected deviations is given by the standard deviations $\sigma_{km}$ so that we need to compute
\begin{align}
    \varepsilon=\sqrt{\sum_{k,m=1}^n \sigma_{km}^2}=\sqrt{\sum_{k=1}^n \sigma_{k}^2+\sum_{k<m} 2\sigma_{km}^2}\ .
\end{align}

\subsubsection{Hessian shift rule}
The variance for a Hessian diagonal entry $H_{kk}$ is $\sigma^2R_k^4/N_{kk}$ if we use $N_{kk}$ shots to estimate it (see Eq.~\eqref{eq:shots_genpar2})\footnote{Recall that $\sigma^2$ is the single-shot variance.}.
For an off-diagonal element $H_{km}$ computed via the diagonal shift rule in Eq.~\eqref{eq:hessian_parshift}, the variance is
\begin{align}
    \sigma_{km}^2=\frac{1}{4}\left(\frac{\sigma^2 (R_k+R_m)^4}{N_{km}}+\frac{\sigma^2R_k^4}{N_{kk}}+\frac{\sigma^2R_m^4}{N_{mm}}\right),
\end{align}
where we used that $R_{km}=R_k+R_m$ for equidistant frequencies. Overall, this yields
\begin{align}
    \varepsilon^2=\sum_{k=1}^n\frac{\sigma^2 R_k^4}{N_{kk}}\frac{n+1}{2}+\sum_{k<m}\frac{\sigma^2(R_k+R_m)^4}{2N_{km}}
\end{align}
If we allocate $N_\text{diag}$ shots optimally, that is $N_{km}$ is proportional to the square root of the coefficient of $N_{km}^{-1}$, we require
\begin{align}
    N_\text{diag} &= \frac{\sigma^2}{\varepsilon^2}\left[\sum_{k=1}^n R_k^2\sqrt{\frac{n+1}{2}}+\sum_{k<m}\frac{1}{\sqrt{2}}(R_k+R_m)^2\right]^2\nonumber\\
    &=\frac{\sigma^2}{2\varepsilon^2}\Big[\bigl(\sqrt{n+1}+n-2\bigr)\norm{\RR}_2^2+\norm{\RR}_1^2\Big]^2
\end{align}
shots to estimate $H$ to a precision of $\varepsilon$.

\subsubsection{Repeated general parameter-shift rule}
Without the diagonal shift rule, we compute $H_{km}$ by executing the univariate general parameter-shift rule in Eq.~\eqref{eq:parshift_eq} for $x_k$ and $x_m$ successively, i.e., we apply the rule for $x_m$ to all terms from the rule for $x_k$.
This leads to $4R_kR_m$ terms with their coefficients arising from the first-order shift rule coefficients by multiplying them together:
\begin{align}
    \norm{\yy^{(km)}}_1&=\frac{1}{4R_kR_m}\sum_{\mu=1}^{R_k}\frac{1}{\sin^2(x_\mu)}\sum_{\mu'=1}^{R_m}\frac{1}{\sin^2(x_{\mu'})}\nonumber\\
    &=R_kR_m,
\end{align}
where we used Eq.~\eqref{eq:parshift_coefficient_norm_1}.
Correspondingly, the variance for $H_{km}$ computed by this methods with an optimal shot allocation of $N_{km}$ shots is $\sigma_{km}^2 = \sigma^2 R_k^2R_m^2/N_{km}$.
The mean square of the Frobenius norm then is
\begin{align}
    \varepsilon^2 = \sum_{k=1}^n\frac{\sigma^2 R_k^4}{N_{kk}}+\sum_{k<m}\frac{2\sigma^2R_k^2R_m^2}{N_{km}}
\end{align}
and an optimal shot allocation across the entries of the Hessian to achieve a precision of $\varepsilon$ will require
\begin{align}
    N_\text{genPS} &= \frac{\sigma^2}{\varepsilon^2}\left[\sum_{k=1}^nR_k^2+\sum_{k<m}\sqrt{2}R_kR_m\right]^2\nonumber\\
    &=\frac{\sigma^2}{2\varepsilon^2}\Big[\bigl(\sqrt{2}-1\bigr)\norm{\RR}_2^2+\norm{\RR}_1^2\Big]^2
\end{align}
shots in total.

\subsubsection{Decomposition and repeated original shift rule}
For the third approach, we only require the observation that again all (unique) Hessian entries are estimated independently and that the coefficients arise from all products of two coefficients from the separate shift rules for $x_k$ and $x_m$. This yields $4\p_k\p_m$ coefficients with magnitude $1/4$, so that the calculation of $\varepsilon$ is the same as for the previous approach, replacing $\RR$ by $\p$.
The required shot budget for a precision of $\varepsilon$ is thus
\begin{align}
    N_\text{decomp} &=\frac{\sigma^2}{2\varepsilon^2}\Big[\bigl(\sqrt{2}-1\bigr)\norm{\pp}_2^2+\norm{\pp}_1^2\Big]^2
\end{align}

\section{Generalization to arbitrary spectra}\label{sec:gen}
Throughout this work, we mostly focused on cost functions $E$ with equidistant --- and thus, by rescaling, integer-valued --- frequencies $\{\Omega_\ell\}$.
Here we will discuss the generalization to arbitrary frequencies, mostly considering the changed cost.

\subsection{Univariate functions}\label{sec:univariate_gen}
The nonuniform DFT used to reconstruct the full function $E$ in Sec.~\ref{sec:full_reconstruction}, and its modifications for the odd and even part in Secs.~\ref{sec:odd_reconstruction} and~\ref{sec:even_reconstruction}, can be used straightforwardly for arbitrary frequencies.
However, choosing equidistant shift angles $\{x_\mu\}$ will no longer make the DFT uniform, as was the case for equidistant frequencies.
Correspondingly, the explicit parameter-shift rules for $E'(0)$ and $E''(0)$ in Eqs.~(\ref{eq:parshift_eq}, \ref{eq:parshift_eq2}) do not apply and in general we do not know a closed-form expression for the DFT or the parameter-shift rules.
Symbolically, the parameter-shift rule takes the form
\begin{align}
    E'(0)&=\sum_{\mu=1}^R y^{(1)}_\mu [E(x_\mu)-E(-x_\mu)]\label{eq:parshift_gen}\\
    E''(0)&=y^{(2)}_0E(0) + \sum_{\mu=1}^R y^{(2)}_\mu [E(x_\mu)+E(-x_\mu)].\label{eq:parshift_gen2}
\end{align}

Regarding the evaluation cost, the odd part and thus odd-order derivatives can be obtained at the same price of $2R$ evaluations of $E$ as before, but the even part might no longer be periodic in general; as a consequence,
\begin{align}
    E_\text{even}(\pi)=\frac{1}{2}(E(\pi)+E(-\pi))\neq E(\pi)
\end{align}
actually may require two evaluations of $E$, leading to $2R+1$ evaluations overall.
If the even part is periodic, which is equivalent to all involved frequencies being commensurable, with some period $T$, evaluating $E_\text{even}(T/2)$ allows to skip the additional evaluation.

When comparing to the first derivative based on a decomposition into $\p$ parametrized elementary gates, the break-even point for the number of unique circuits remains at $R=\p$ as for equidistant frequencies, but we note that e.g., a decomposition of the form
\begin{align}\label{eq:decomposition_with_prefactors}
    U(x)=\prod_{k=1}^\p U_k(\beta_k x),
\end{align}
namely where $x$ is rescaled individually in each elementary gate by some $\beta_k\in\R$,
in general will result in $R=\p^2$ frequencies of $E$, making the decomposition-based parameter-shift rule beneficial.
For the second-order derivative, the number of evaluations $2R+1$ might be quadratic in $\p$ in the same way, but the decomposition requires $2\p^2-\p+1$ as well, so that the requirements are similar if $R=\p$.

Regarding the required number of shots, we cannot make concrete statements for the general case as we don't have a closed-form expression for the coefficients $\yy$, but note that for the decomposition approach, rescaling factors like the $\{\beta_k\}$ in Eq.~\eqref{eq:decomposition_with_prefactors} above have to be factored in via the chain rule, leading to a modified shot requirement.

An example for unitaries with non-equidistant frequencies would be the QAOA layer that implements the time evolution under the problem Hamiltonian (see Eq.~\eqref{eq:qaoa_maxcut}) for $\maxcut$ on \emph{weighted} graphs with non-integer weights.

For the stochastic parameter-shift rule in Sec.~\ref{sec:stoch_parshift} we did not restrict ourselves to equidistant frequencies and derive it in App.~\ref{sec:derivation_stoch_parshift} for general unitaries of the form $U_F=\exp(i(xG+F))$ directly.

\subsection{Multivariate functions}\label{sec:multivariate_gen}
While the univariate functions do not differ strongly for equidistant and arbitrary frequencies in $E$ and mostly the expected relation between $R$ and $\p$ changes, the shift rule for the Hessian and the metric tensor are affected heavily by generalizing the spectrum.
First, the univariate restriction $E^{(km)}(x)$ in Eq.~\eqref{eq:simul_shift} still can be used to compute the off-diagonal entry $H_{km}$ of the Hessian but this may require up to $2R_{km}+1=4R_kR_m+2R_k+2R_m-3$ evaluations (see App.~\ref{sec:derivation_hessian_rule}), in contrast to $2R_{km}=2(R_k+R_m)$ in the equidistant case.
Compared to the resource requirements of the decomposition-based approach, $4\p_k\p_m$, this makes our general parameter-shift rule more expensive if $R_k\gtrsim \p_k$.

As we use the same method to obtain the metric tensor $\mathcal{F}$, the number of evaluations grows in the same manner, making the decomposition-based shift rule more feasible for unitaries with non-equidistant frequencies.
As $f(\xx_0)$ does not have to be evaluated, an off-diagonal element $\mathcal{F}_{km}$ requires one evaluation fewer than $H_{km}$, namely $4R_kR_m+2R_k+2R_m-4$.

\section{General stochastic shift rule}\label{sec:derivation_stoch_parshift}
In this section we describe a stochastic variant of the general parameter-shift rule which follows immediately from combining the rule for single-parameter gates in Eq.~\eqref{eq:parshift_gen} with the result from Ref.~\cite{stoch_parshift}.

First, note that any shift rule 
\begin{align}
    E'(x_0)=\sum_\mu y_\mu E(x_0+x_\mu),
\end{align}
with coefficients $\{y_\mu\}$ and shift angles $\{x_\mu\}$ for a unitary $U(x)=\exp(ixG)$, implies that we can implement the commutator with $G$:
\begin{align}\label{eq:commutator_via_parshift}
    i[G,\rho] = \sum_\mu y_\mu U(x_\mu) \rho U^\dagger(x_\mu),
\end{align}
since the commutator between $G$ and the Hamiltonian directly expresses the derivative of the expectation value $E'(0)$ on the operator level, and shift rules hold for arbitrary states.

Now consider the extension $U_F(x)=\exp(i(xG+F))$ of the above unitary.
In the original stochastic parameter-shift rule, the authors show\footnote{To be precise, we here combine Eqs.~(11-13) in Ref.~\cite{stoch_parshift} into a general expression for $E'$.}
\begin{align}
    E'(x_0)=&\int_0^1 \mathrm{d}t \;\operatorname{tr}\bigg\{U_F^\dagger(tx_0)B\,U_F(tx_0) \\
    &\hspace{-0.6cm}\times i\left[G\ ,\  U_F\bigl((1-t)x_0\bigr)\ket{\psi}\!\!\bra{\psi}U_F^\dagger\bigl((1-t)x_0\bigr)\right]\bigg\}\nonumber
\end{align}
where we again denoted the state prepared by the circuit before $U_F$ by $\ket{\psi}$ and the observable transformed by the circuit following $U_F$ by $B$.
By using Eq.~\eqref{eq:commutator_via_parshift} to express the commutator, we obtain
\begin{align}\label{eq:_stoch_parshift_intermed}
    E'(x_0)&=\int_0^1 \mathrm{d}t \;\sum_\mu y_\mu \operatorname{tr}\bigg\{U_F^\dagger(tx_0)B\,U_F(tx_0) \\
        &\hspace{-1.2cm}\times U(x_\mu) U_F\bigl((1-t)x_0\bigr)\ket{\psi}\!\!\bra{\psi}U_F^\dagger\bigl((1-t)x_0\bigr)U^\dagger(x_\mu)\bigg\}.\nonumber
\end{align}
We abbreviate the interleaved unitaries 
\begin{align}
    U_{F,\mu}(x_0, t)\coloneqq U_F(tx_0)U(x_\mu)U_F\bigl((1-t)x_0\bigr)
\end{align}
and denote the cost function that uses $U_{F,\mu}(x_0,t)$ instead of $U_F(x_0)$ as
\begin{align}
    E_\mu(x_0, t) \coloneqq \operatorname{tr}\left\{ B\  U_{F,\mu}^\dagger(x_0,t)\ket{\psi}\!\!\bra{\psi}U_{F,\mu}(x_0,t) \right\}.\nonumber
\end{align}
Rewriting Eq.~\eqref{eq:_stoch_parshift_intermed} then yields the \emph{generalized stochastic parameter-shift rule}
\begin{align}
    E'(x_0)=\int_0^1\mathrm{d}t\sum_\mu y_\mu E_\mu(x_0, t).
\end{align}
It can be implemented by sampling values for the splitting time $t$, combining the shifted energies $E_\mu(x_0,t)$ for each sampled $t$ with the coefficients $y_\mu$, and averaging over the results.

\section{Details on QAD}\label{sec:derivation_qad}
In this section we provide details on the latter two of the three modifications of the QAD algorithm discussed in Sec.~\ref{sec:qad}.

\subsection{Extended QAD model for Pauli rotations}\label{sec:derivation_extQAD}
The QAD model introduced in Ref.~\cite{qad} contains trigonometric functions up to second (leading) order.
The free parameters of the model cannot be extracted with one function evaluation per degree of freedom, because unlike standard monomials in a Taylor expansion, the trigonometric basis functions mix the orders in the input parameters.
This leads to the mismatch of $2n^2+n+1$ (original QAD) or $3n^2/2+n/2+1$ (see above) evaluations to obtain $n^2/2+3n/2+1$ model parameters.
We note that the QAD model contains full univariate reconstructions at optimal cost, extracting the $2n+1$ model parameters $E^{(A)}$, $\EE^{(B)}$ and $\EE^{(C)}$ from $2n+1$ function evaluations.
The doubly shifted evaluations, however, are used for the Hessian entry only:
\begin{align}
    E^{(D)}_{km}=\frac{1}{4}\left[E^{++}_{km}-E^{+-}_{km}-E^{-+}_{km}+E^{--}_{km}\right],
\end{align}
where $E^{\pm\pm}_{km}=E(\xx_0\pm\frac{\pi}{2}\vv_k\pm\frac{\pi}{2}\vv_m)$ and we recall that this QAD model is restricted to Pauli rotations only. 

Let us now consider a slightly larger truncation of the cost function than the one presented in App.~A~2 in \cite{qad}:
\begin{align}\label{eq:extQAD_funcform}
    \mathring{E}(\xx_0+\xx)&=A(\xx)\biggl[E^{(A)}\nonumber\\
    &\hspace{-1.4cm}+2\EE^{(B)}\cdot \tan\left(\frac{\xx}{2}\right)+2\EE^{(C)}\cdot \tan\left(\frac{\xx}{2}\right)^{\odot 2}\nonumber\\
    &\hspace{-1.4cm}+4\tan\left(\frac{\xx}{2}\right)
    E^{(D)} \tan\left(\frac{\xx}{2}\right) \\
    &\hspace{-1.4cm}+4\tan\left(\frac{\xx}{2}\right)
    E^{(F)} \tan^2\left(\frac{\xx}{2}\right)\nonumber\\
    &\hspace{-1.4cm}+4\tan^2\left(\frac{\xx}{2}\right)
    E^{(G)} \tan^2\left(\frac{\xx}{2}\right)\biggr]\nonumber
\end{align}
with $A(\xx)=\prod_k \cos^2(x_k/2)$.
$E^{(F)}$ and $E^{(G)}$ have zeros on their diagonals because there are no terms of the form $\sin^3(x_k/2)$ or $\sin^4(x_k/2)$ in the cost function, and for $E^{(G)}$ we only require the strictly upper triangular entries due to symmetry.
The higher-order terms contain at least three distinct variables $x_k$, $x_l$ and $x_m$ because all bivariate terms are captured in the above truncation.
Using
\begin{align}
    A\left(\pm\frac{\pi}{4}\vv_k\pm\frac{\pi}{4}\vv_m\right)=\frac{1}{4} \;\text{ and }\; \tan\left(\pm\frac{\pi}{4}\right)=\pm1,\nonumber
\end{align}
we now can compute:
\begin{align*}
    E^{++}_{km}-E^{-+}_{km}+E^{+-}_{km}-E^{--}_{km} &= E^{(B)}_k+E^{(F)}_{km}\\
    E^{++}_{km}+E^{-+}_{km}+E^{+-}_{km}+E^{--}_{km} &= E^{(A)}+2E^{(C)}_k\\
                                                    &+2E^{(C)}_m+4E^{(G)}_{km}.
\end{align*}
This means that the $4$ function evaluations $E^{\pm\pm}_{km}$ that are used for $E_{km}^{(D)}$ in the original QAD can be recycled to obtain the $3$ parameters $E_{km}^{(F)}$, $E_{mk}^{(F)}$ and $E_{km}^{(G)}$.
The corresponding model is of the form Eq.~\eqref{eq:extQAD_funcform} and therefore includes \emph{all} terms that depend on two parameters only.
Consequentially, the constructed model exactly reproduces the cost function not only on the coordinate axes but also on all coordinate planes spanned by any two of the axes.
The number of model parameters is $2n^2+1$, which matches the total number of function evaluations.

\subsection{Trigonometric interpolation for QAD}\label{sec:derivation_genQAD}
Both the original QAD algorithm, and the extension introduced above, assume the parametrized quantum circuit to consist of Pauli rotation gates exclusively.
In the spirit of the generalized function reconstruction and parameter-shift rule, we would like to relax this assumption and generalize the QAD model.
However, there is no obvious unique way to do this, because the correspondence between the gradient and $\EE^{(B)}$ and between the Hessian and $\EE^{(C,D)}$ is not preserved for multiple frequencies.
Instead, the uni- and bivariate Fourier coefficients of $E$ form the model parameters and the derivative quantities are contractions with the frequencies thereof.
There are multiple ways in which we could generalize QAD to multiple frequencies.

The first way to generalize QAD is to compute the gradient and Hessian with the generalized parameter-shift rule Eq.~\eqref{eq:parshift_eq} and the shift rule for Hessian entries Eq.~\eqref{eq:hessian_parshift} and to construct a single-frequency model as in original QAD.
Even though we know the original energy function to contain multiple frequencies, this would yield a local model with the correct second-order expansion at $\xx_0$ that exploits the evaluations savings shown in this work.
As QAD is supposed to use the model only in the neighbourhood of $\xx_0$, this might be sufficient for the optimization.

As a second generalization we propose a full trigonometric interpolation of $E$ up to second order, similar to the univariate reconstruction in Sec.~\ref{sec:full_reconstruction}.
First we consider the univariate part of the model:
Start by evaluating $E$ at positions shifted in the $k$th coordinate by equidistant points and subtract $E(\xx_0)$,
\begin{align}
    E_\mu^{(k)} &\coloneqq E(\xx_0+x_\mu\vv_k)-E(\xx_0)\\
    x_\mu &\coloneqq \frac{2\mu\pi}{2R_k+1}, \quad \mu\in[2R_k].
\end{align}
Then consider the (shifted) Dirichlet kernels
\begin{align}
    \D_\mu^{(k)}(x)&=\frac{1}{2R_k+1}\left(1+2\sum_{\ell=1}^{R_k} \cos(\ell (x-x_\mu)) \right)\\
    &=\frac{\sin\left(\frac{1}{2}(2R_k+1)(x-x_\mu)\right)}{(2R_k+1)\sin\left(\frac{1}{2}(x-x_\mu)\right)}
\end{align}
which satisfy $\D^{(k)}_\mu(x_{\mu'})=\delta_{\mu\mu'}$ and are Fourier series with integer frequencies up to $R_k$.
Therefore, the function\footnote{One might be wondering why to subtract $E(\xx_0)$ just to add it manually back into the reconstruction now. This is because we need to avoid duplicating this term when adding up the univariate and bivariate terms of all parameters later on.}
\begin{align}
    \hat{E}^{(k)}(x)=\sum_{\mu=1}^{2R_k} E_\mu^{(k)}\D^{(k)}_\mu(x)
\end{align}
coincides with $E(\xx_0+x\vv_k)-E(\xx_0)$ at $2R_k+1$ points and is a trigonometric polynomial with the same $R_k$ frequencies.

Similarly, the product kernels $\D_{\mu\mu'}^{(km)}(x_k, x_m)=\D_\mu^{(k)}(x_k)\D_{\mu'}^{(m)}(x_m)$ can be used to reconstruct the bivariate restriction of $E$ to the $x_k-x_m$ plane.
For this, evaluate the function at doubly shifted positions and subtract both, $E(\xx_0)$ and the univariate parts:
\begin{align}
    E_{\mu\mu'}^{(km)} &\coloneqq E(\xx_0+x_\mu\vv_k+x_{\mu'}\vv_m)\\
    &-\hat{E}^{(k)}(x_\mu)-\hat{E}^{(m)}(x_{\mu'})-E(\xx_0)
\end{align}
Then, the bivariate Fourier series
\begin{align}
    \hat{E}^{(km)}(x_k, x_m)=\sum_{\mu,\mu'=1}^{2R_k,2R_m} E_{\mu\mu'}^{(km)}\D_{\mu\mu'}^{(km)}(x_k, x_m)
\end{align}
coincides with $E(\xx_0+x_k\vv_k+x_m\vv_m)-E(\xx_0)-\hat{E}^{(k)}(x_k)-\hat{E}^{(m)}(x_m)$ on the entire coordinate plane spanned by $\vv_k$ and $\vv_m$.

As we constructed the terms such that they do not contain the respective lower order terms, we finally can combine them to the full trigonometric interpolation:
\begin{align}
    \hat{E}_\text{interp}(\xx)=E(\xx_0)&+\sum_{k=1}^n \hat{E}^{(k)}(x_k)\\
    &+\sum_{k<m} \hat{E}^{(km)}(x_k,x_m).\nonumber
\end{align}
This model has as many parameters as function evaluations, namely $2(\norm{\RR}_1^2-\norm{\RR}_2^2+\norm{\RR}_1)+1$, and therefore, the trigonometric interpolation is the generalization of the extended QAD model in App.~\ref{sec:derivation_extQAD}.
Indeed, for $R_k=1$ for all $k$ we get back $2(n^2-n+n)+1=2n^2+1$ evaluations and model parameters.

We note that the trigonometric interpolation can be implemented for non-equidistant evaluation points in a similar manner and with the same number of evaluations, although the elementary functions are no longer Dirichlet kernels but take the form
\begin{align}
    \Dnoneq_\mu^{(k)}(x)=\frac{\sin\left(\frac{1}{2}x\right)}{\sin\left(\frac{1}{2}x_\mu\right)}\prod_{\mu'=1}^{2R_k}\frac{\sin\left(\frac{1}{2}(x-x_{\mu'})\right)}{\sin\left(\frac{1}{2}(x_\mu-x_{\mu'})\right)}.
\end{align}
\end{appendix}
\end{document}